\documentclass[prd,twocolumn,superscriptaddress,floatfix,amsmath,amssymb,amsfonts,nofootinbib,longbibliography]{revtex4-2}

\usepackage{float} 
\usepackage{scalerel}
\usepackage[normalem]{ulem}
\usepackage[english]{babel}
\usepackage{graphicx}
\usepackage{dcolumn}
\usepackage{bm}
\usepackage{blindtext}
\usepackage{verbatim}
\usepackage{relsize}
\usepackage{mathrsfs}
\usepackage{musicography}
\usepackage{amsmath}
\usepackage{blindtext}
\usepackage{cancel}
\usepackage{physics}
\usepackage{epstopdf}
\usepackage{mathtools}
\usepackage{blindtext}
\usepackage{tensor}
\usepackage{color}
\usepackage[usenames,dvipsnames]{pstricks}
\usepackage{epsfig}
\usepackage{pst-grad} 
\usepackage{pst-plot} 
\usepackage{hyperref}
\usepackage{verbatim}
\usepackage{slashed}
\usepackage{dsfont}

\hypersetup{
    colorlinks = true,
    linkcolor = blue,
    anchorcolor = blue,
    citecolor = blue,
    urlcolor = cyan
    }

\newcommand{\mf}{\mathsf}

\newcommand{\ii}{\mathrm{i}}

\newcommand{\tc}[1]{\textsc{#1}}

\allowdisplaybreaks[1] 

\begin{document}

\title{Particle detector models from path integrals of localized quantum fields}

\author{Bruno de S. L. Torres}
\email{bdesouzaleaotorres@perimeterinstitute.ca}\affiliation{Perimeter Institute for Theoretical Physics, Waterloo, Ontario, N2L 2Y5, Canada}
\affiliation{Institute for Quantum Computing, University of Waterloo, Waterloo, Ontario, N2L 3G1, Canada}
\affiliation{Department of Physics and Astronomy, University of Waterloo, Waterloo, Ontario, N2L 3G1, Canada}

\begin{abstract}
    Using the Schwinger-Keldysh path integral, we draw a connection between localized quantum field theories and more commonly used models of local probes  in Relativistic Quantum Information (RQI). By integrating over and then tracing out the inaccessible modes of the localized field being used as a probe, we show that, at leading order in perturbation theory, the dynamics of any finite number of modes of the probe field is exactly that of a finite number of harmonic-oscillator Unruh-DeWitt (UDW) detectors. The equivalence is valid for a rather general class of input states of the probe-target field system, as well as for any arbitrary number of modes included as detectors. The path integral also provides a closed-form expression which gives us a systematic way of obtaining the corrections to the UDW model at higher orders in perturbation theory due to the existence of the additional modes that have been traced out. This approach vindicates and extends a recently proposed bridge between detector-based and field-theory-based measurement frameworks for quantum field theory~\cite{QFTPD}, and also points to potential connections between particle detector models in RQI and other areas of physics where path integral methods are more commonplace---in particular, the Wilsonian approach to the renormalization group and effective field theories.
\end{abstract}

\maketitle

\section{Introduction}
Our modern understanding of the elementary building blocks of the physical world is ultimately based on the concept of \emph{fields}. General relativity---currently our most successful theory of gravity---is fundamentally a theory about the dynamics of the gravitational field and how it manifests as the geometry of spacetime. Similarly, the Standard Model of particle physics---our most complete description of the fundamental constituents of matter and all elementary forces of nature (excluding gravity)---is entirely formulated in the language of quantum field theory (QFT), where the concept of quantum field is essential to combine the principles of quantum mechanics and special relativity into a single, consistent framework. This appears to be one of the most profound lessons of modern theoretical physics.

It also appears to be a deep truth that the only way we can acquire information about fields is through some form of interaction between the field of interest and an auxiliary system that can couple to it in localized regions of space and time. We learn about the structure of the gravitational field by studying how it affects the trajectories of (approximately) pointlike test particles; we learn about the electromagnetic field by tracking the dynamics of localized charges and currents; and although all the elementary constituents of matter ultimately emerge from relativistic quantum fields themselves, our knowledge about their most fundamental properties ultimately comes from the localized excitations detected by the countless sensors that make up particle accelerators such as the LHC---which, directly or indirectly, carries information about the intricate way those fields interact. It should therefore come as no surprise that a consistent framework for how to extract and process information from quantum fields will in general be intimately tied to how local probes can interact with quantum fields. 

This general principle is embodied most concretely in the context of Relativistic Quantum Information (RQI). This is an area of research---mostly focused on exploring information-theoretic features of quantum field theory and gravity, as well as studying the role that relativity may play in information-processing tasks---where local probes (usually referred to as \emph{particle detectors}) play a central role. Particle detectors in RQI consist of idealized versions of localized quantum systems that can controllably couple to quantum fields in local regions of spacetime. From a theoretical perspective, these have proven to be very useful for investigations on several aspects of the interplay between quantum information, quantum field theory, and gravity: applications range from a measurement framework for QFT which can be made to respect the underlying principles of relativistic locality and causality~\cite{jose}, operationally motivated protocols for quantifying and extracting entanglement from QFTs~\cite{Reznik2003, Reznik1, Pozas-Kerstjens:2015, kelly}, and detector-based approaches to important phenomena at the foundations of QFT in both flat and curved spacetimes, such as the Unruh effect and Hawking radiation~\cite{Unruh1976, Sciama1977, DeWitt}. From a more practical point of view, particle detectors also capture important features of physical setups of great experimental significance, such as the dynamics of atomic probes coupled to the electromagnetic field in atomic physics and quantum optics~\cite{richard}. They are, therefore, very well-adapted as theoretical tools for the study of relativistic information-processing tasks that can (at least in principle) actually be performed in a lab.

For practical purposes, it is common to assume that the internal dynamics of the system being used as a detector can be approximated as non-relativistic. This is a natural assumption to make when one thinks of archetypical examples of particle detectors given, for instance, by atomic probes coupled to external electromagnetic fields: after all, in many regimes of physical interest, the internal structure of the atom can be well-approximated by the non-relativistic Schrodinger equation applied to the electrons under the influence of the atomic nucleus. 

Approximating the probe as an internally non-relativistic system is also advantageous from a purely theoretical point of view. It is well-known that building a framework for local measurements entirely within relativistic QFT is highly nontrivial, since most immediate versions of measurement postulates imported from non-relativistic quantum mechanics often lead to irreconcilable conflicts with relativistic locality and causality~\cite{Sorkin, dowker2011, impossibleRevisited}. In contrast, if we can approximate the probe's internal dynamics as non-relativistic, we are justified in applying the standard measurement framework from non-relativistic quantum mechanics to the physical system being used as a detector. This dramatically alleviates the burden of directly measuring a relativistic field, by replacing this problem with the (arguably easier) problem of measuring the detectors, and then indirectly inferring properties of the field by letting the detectors couple to it. 

However, there is something slightly vexing about ending the story here. After all, our most fundamental theory for the structure of matter teaches us that the internal constituents of any physical system---including the probe itself---fundamentally emerge from quantum fields. This has brought a lot of attention to the question of how to model measurements in quantum field theory using fully relativistic probes. Progress in this direction is spearheaded by the Fewster-Verch (FV) framework~\cite{fewster1, fewster2}, which is an approach to the measurement problem in QFT where the probe system is also treated as a fully relativistic field theory, formulated in the language of algebraic QFT in general (possibly curved) spacetimes. 

The FV framework has been remarkably successful at providing a mathematically rigorous measurement scheme for QFT that is fully compatible with the relativistic nature of the theory. In particular, the formalism provides an elegant solution to long-standing problems regarding how to make sense of local measurements on quantum fields while taking relativistic locality and causality into account~\cite{impossibleImpossible, singularOrNonlocal}.\footnote{We should emphasize that the detector-based approach of~\cite{jose} is also capable of addressing these problems; this was, in fact, part of the motivation for its inception. However, the approximation of the probe as a nonrelativistic system comes at a price. For more details, see e.g.~\cite{PipoFTL}---or wait until Subsec.~\ref{sub:covcausality}.} However, since the probe in this case is also a quantum field, the FV framework is agnostic to how an external agent (say, an experimentalist taking notes on their note pad) can in practice extract information from their local measurements. Strictly speaking, the framework only describes the chain of information flow from one relativistic quantum field (the target field we are actually interested in measuring) to another relativistic quantum field (the probe system being used as a detector). It is, therefore, less directly applicable to more realistic measurement settings, especially when compared to the detector-based framework of~\cite{jose}.

Our goal in this paper is to show how to connect a simple, fully relativistic model of local probe---namely, a scalar field that is effectively confined to a localized region of space by an external potential---to a system of local probes that is closer in spirit to the usual particle detector models adopted in RQI. This is an attempt towards a more comprehensive understanding of the relation between particle detector models and field-theoretic descriptions of local probes, with the aim of narrowing the gap between the FV framework and the detector-based approaches to RQI.\footnote{On a related note, we also draw attention to recent accounts~\cite{Dan2023, MariaDoreen2023} which thoroughly discuss how particle detector models and the FV framework both fit within the bigger picture of measurements in QFT, seen from the point of view of foundations and philosophy of physics.}

A first step in this direction was taken in recent work~\cite{QFTPD} using standard techniques from particle detector calculations in RQI---in particular, studying the dynamics of quantum field and probe by taking the leading-order terms in the Dyson expansion of the time evolution operator, and then evolving the initial state of the system in the interaction picture. Here, however, we will adopt a somewhat complementary approach, taking full advantage of path integral methods to describe the dynamics of the field and the probe. This will allow us to forego any explicit use of perturbation theory, and recover the results of~\cite{QFTPD} as a particular case of a much more general statement relating the field-theoretic and detector-based descriptions of probes for quantum fields.

The main technical tool employed here will be the \emph{Schwinger-Keldysh path integral}~\cite{SchwingerPathIntegral, KeldyshPathIntegral, FeynmanVernon}. This can be thought of as an upgraded version of the Feynman path integral that is better suited for describing the non-equilibrium dynamics of a system interacting with an inaccessible environment. In this case, the role of the ``environment'' is played, in a loose sense, by a set of degrees of freedom of the probe system that are deemed inaccessible, and are therefore traced out in the final state of the system of interest. The result is an effective description of the detector-field system that only contains finitely many degrees of freedom of the probe---which resembles what would be obtained if one started with a finite number of detectors coupled to the field. Due to the simplicity of the model for the probe, the effect of integrating over the inaccessible degrees of freedom can be worked out in full detail. This therefore provides a simple setting in which to compare predictions from particle detector models and more field-theoretic descriptions for probe systems.

The Schwinger-Keldysh path integral has already been suggested as an important tool for the development of field-theory-based formulations of RQI~\cite{charis2023, charis2023Again}. The way it is being used here, however, highlights its promising role as a method of explicitly connecting fully relativistic probes to effectively nonrelativistic particle detectors, which is something that has not been proposed before (to the best of the author's knowledge at the time of writing). This paves the way for further investigations on the interplay between measurement schemes in RQI and techniques from effective field theory~\cite{polchinski1999effective, BurgessEffectiveFieldTheory}---which, at least on a conceptual level, is clearly the correct framework in which to address the apparent tension between fundamentally relativistic dynamics of the probes and nonrelativistic descriptions of measurements in experimentally relevant scales.

The paper is organized as follows. Section~\ref{sec:detectors} reviews the basics of particle detector models most commonly used in RQI---in particular the Unruh-DeWitt (UDW) model---, and briefly discusses some of the shortcomings of these models when taking into account what one would expect from a fully relativistic theory. In Section~\ref{sec:LocalQF} we describe a simple model of a fully relativistic probe, consisting of scalar field that is effectively localized by some external potential. This will be used as the concrete model for a local probe that remedies the main issues raised at the end of Sec.~\ref{sec:detectors}. We then move on to Section~\ref{sec:PathIntegral}, where we describe the dynamics of the joint field-detector system via the Schwinger-Keldysh path integral, and show how to effectively reduce the localized quantum field to a finite number of UDW detectors by explicitly integrating over and tracing out a set of inaccessible degrees of freedom of the probe. The remaining path integral for the modes that are kept as detector degrees of freedom is, at leading order in perturbation theory, precisely the same as what would be obtained in the UDW model; at higher orders, the path integral also provides a closed-form expression which can be used to systematically calculate the deviations between the two models. Section~\ref{sec:multipleminima} contains a simple extension of the framework described in Sec.~\ref{sec:PathIntegral}, describing how detectors in multiple trajectories can emerge from the same underlying localized quantum field. We also comment on a few examples where this simple extension may be of physical relevance. In Section~\ref{sec:conclusions} we summarize our work, and comment on a few possible directions for future research.

\textbf{Notation and conventions}. Spacetime is given by a pair $(\mathcal{M}, g_{ab})$, where $\mathcal{M}$ is a $(d+1)$-dimensional differentiable manifold and $g_{ab}$ is a Lorentzian metric on $\mathcal{M}$. For simplicity, it will always be assumed that the background spacetime in question is globally hyperbolic. The signature convention for the metric is such that $g_{ab}T^a T^b <0$ if $T^a$ is a timelike vector. We will represent abstract points in $\mathcal{M}$ with sans-serif font $\mf{x}$, and reserve the normal math font $x$ for the collection of spacetime coordinates associated to the point $\mf{x}$ in a given coordinate system. In Section~\ref{sec:detectors}, $(t, \bm{x})$ will denote any set of coordinates such that $t$ is a timelike coordinate (i.e., the $1$-form $\dd t$ is such that $g^{ab}\left(\dd t\right)_a\left(\dd t\right)_b < 0$), and $\bm{x} = (x^1, \dots, x^d)$ are spatial coordinates in the surfaces of constant $t$; from Section~\ref{sec:LocalQF} onwards, however, $(t, \bm{x})$ will exclusively refer to static coordinates in $\mathcal{M}$, for which the metric takes the form~\eqref{eq:easymetric}. We will denote by $\nabla_a$ the Levi-Civita connection, defined as the unique torsion-free covariant derivative that is compatible with the metric $g_{ab}$. $\dd V$ is the volume form associated to $g_{ab}$, which in any coordinate system takes the form $\dd V = \sqrt{-g}\,\dd^{d+1}x$, where $g\equiv \det(g_{\mu\nu})$. Quantum operators acting on a Hilbert space will always be written with hats, to clearly distinguish them from their classical counterparts: for instance, $\hat{\phi}_{\tc{d}}(t, \bm{x})$ in Eq.~\eqref{eq:probefieldeigenstate} is the operator (or, if you want to be more pedantic, the operator-valued distribution) corresponding to the quantized version of $\phi_{\tc{d}}(t, \bm{x})$---which, when written as in Eq.~\eqref{eq:modesum}, for example, is just a classical field configuration.  We adopt natural units, with $\hbar = c = 1$.  
\section{Particle Detector models}\label{sec:detectors}
In this section we review some of the general aspects of particle detector models, with the Unruh-DeWitt (UDW) model being introduced as a paradigmatic example. We also discuss some of the drawbacks and limitations of particle detector models---especially pertaining to considerations about covariance and causality---which have been used to advocate for the necessity of a fully relativistic, field-theoretic version of local probes for quantum fields.
\subsection{General lore and UDW model}\label{sub:GenPD}
The general philosophy when using particle detectors as probes for quantum fields is to extract information from the quantum field of interest by coupling a detector to it, and studying how the evolution of the state of the detector indirectly depends on features of the quantum field being probed. In very general terms, the dynamics of the joint detector-field system can be described through an action of the form
\begin{equation}\label{eq:genaction}
    S = S_{\tc{f}} + S_{\tc{d}} + S_{\tc{i}},
\end{equation}
where $S_{\tc{f}}$ and $S_{\tc{d}}$ provide the free dynamics of field and detector respectively, and $S_{\tc{i}}$ encodes the coupling between the two systems. 

The action of the field is typically expressed as an integral of a Lagrangian over spacetime,
\begin{equation}
    S_{\tc{f}} = \int \dd V \mathcal{L}_{\tc{f}}(\psi^A, \nabla_a\psi^A, \dots),
\end{equation}
where the Lagrangian $\mathcal{L}_{\tc{f}}$ is a scalar function of the dynamical fields $\psi^A$ and its covariant derivatives, and the index $A$ encodes any possible collection of tensor/spinor indices for the dyamical fields in question. $\mathcal{L}_{\tc{f}}$ also generally depends on some external background fields (such as the spacetime metric $g_{ab}$ and any additional external potentials that are treated as non-dynamical), and may also include some explicit dependence on higher derivatives of the metric through terms such as non-minimal coupling to curvature.

In its simplest version, the detector is pictured as a localized system following a fixed classical trajectory $\mf{z}(\tau)$ on $\mathcal{M}$, where $\tau$ has the interpretation of the detector's proper time. The system is also endowed with a set of quantum internal degrees of freedom $q^i$, whose dynamics can be effectively prescribed by an action which takes the form of an integral over proper time of some Lagrangian,
\begin{equation}
    S_{\tc{d}} = \int\dd\tau \,\mathcal{L}_{\tc{d}}(q^i, \frac{\dd q^i}{\dd\tau}, \dots).
\end{equation}
Similar to $\mathcal{L}_{\tc{f}}$, the detector's Lagrangian $\mathcal{L}_{\tc{d}}$ is a function of the system's internal degrees of freedom and its derivatives, and it will also generically depend on features of the extrinsic geometry of how $\mf{z}(\tau)$ is embedded in $\mathcal{M}$ (for instance, the proper acceleration of the detector's trajectory), as well as other geometrical aspects of the background spacetime (for instance, the Riemann curvature tensor of the metric $g_{ab}$ evaluated along $\mf{z}(\tau)$). There can also be extensions of the formalism where the detector is put to evolve in a \emph{superposition} of trajectories~\cite{SuperTrajectories, UnruhSuperposition}; this can be achieved dynamically, for instance, by letting the detector's centre of mass also be a quantum degree of freedom~\cite{NadineDelocalization, FlaminiaAchim, EvanDelocalization}. For the most part, the simpler setting where the detector's position is treated as classical will be enough for our purposes in this paper. However, we shall come back to briefly comment on a similar extension to detectors associated to more than one classical trajectory in Sec.~\ref{sec:multipleminima}.

Finally, the coupling between detector and field can be expressed through an interaction action of the form
\begin{equation}\label{eq:UDWCoupling}
    S_{\tc{i}} = \lambda \int \dd V \Lambda(\mf{x})\mu_A(\tau(\mf{x}))\mathcal{O}^A(\mf{x}),
\end{equation}
where $\mu_A(\tau)$ represents an observable of the detector, $\mathcal{O}^A(\mf{x})$ is an observable of the field, $\Lambda(\mf{x})$ is a spacetime smearing function that dictates the strength of the interaction between probe and field in space and time, and $\lambda$ is an overall coupling constant.\footnote{When writing Eq.~\eqref{eq:UDWCoupling}, we are implicitly assuming that $\Lambda(\mf x)$, $\mu_A(\tau)$, and $\mathcal{O}^A(\mf{x})$ are all real, and that upon quantization both $\hat{\mu}_A(\tau)$ and $\hat{\mathcal{O}}^A(\mf{x})$ are Hermitian observables on the respective Hilbert spaces of detector and field. An obvious---and most importantly, sometimes physically motivated~\cite{Pozas2016, neutrinos, boris}---generalization of this setup would allow for $\Lambda(\mf x)$ to be complex, and $\hat{\mu}_A(\tau)$ and $\hat{\mathcal{O}}^A(\mf{x})$ to be non-Hermitian; in this case, the interaction action between detector and field would contain a sum of terms which include the right-hand side of~\eqref{eq:UDWCoupling} and its complex conjugate.} The spacetime smearing function $\Lambda(\mf{x})$ should be pictured as most strongly supported around the detector's trajectory $\mf{z}(\tau)$. In Eq.~\eqref{eq:UDWCoupling}, $\tau(\mf{x})$ corresponds to the Fermi normal coordinate time relative to the trajectory $\mf{z}(\tau)$, which extends the proper time parameter $\tau$ (originally only defined for points along the detector's trajectory) to a timelike coordinate that can be assigned to any point $\mf{x}$ in a sufficiently small neighbourhood of $\mf{z}(\tau)$~\cite{poisson}. Once again, the index $A$ in the field observable $\mathcal{O}^A(\mf x)$ can comprise any possible collection of tensorial/spinorial indices, with the detector's observable $\mu_{A}(\tau)$ then being an element of the dual space to $\mathcal{O}^A$ in order for the contraction $\mu_A\mathcal{O}^A$ to form a Lorentz scalar.

Given a coordinate system $(t, \bm{x})$ which foliates $\mathcal{M}$ by a family of Cauchy surfaces $\mathcal{E}_t$ labeled by constant values of the timelike coordinate $t$, the quantum dynamics of the joint system of detector and field interacting through the action~\eqref{eq:UDWCoupling} can also be described in terms of an interaction Hamiltonian
\begin{align}\label{eq:interactionHamiltonian}
    \hat{H}_{\tc{i}}(t) &= -\lambda\int_{\mathcal{E}_t}\dd^d\bm{x}\,\sqrt{-g}\,\Lambda(\mf x)\hat{\mu}_A(\tau(\mf{x}))\,\hat{\mathcal{O}}^A(\mf{x})\\
    &= \int_{\mathcal{E}_t}\dd^d \bm{x}\sqrt{-g}\,\hat{h}_{\tc{i}}(\mf x),\nonumber
\end{align}
where we have conveniently defined the Hamiltonian scalar density
\begin{equation}\label{eq:HDensity}
    \hat{h}_{\tc{i}}(\mf x) = -\lambda \Lambda(\mf x)\,\hat{\mu}_A(\tau(\mf{x}))\,\hat{\mathcal{O}}^A(\mf{x}).
\end{equation}
This is the most common starting point in concrete applications of particle detectors in RQI. In this context, it is also customary to adopt the interaction picture, in which case the operators $\hat{\mu}_A(\tau(\mf{x}))$ and $\hat{\mathcal{O}}^A(\mf{x})$ appearing in~\eqref{eq:interactionHamiltonian} should be understood as the resulting time evolution implied by the free actions $S_{\tc{d}}$ and $S_{\tc{f}}$, respectively. In the interaction picture, the joint state of the detector-field system then evolves according to the unitary time evolution operator
\begin{equation}\label{eq:intimevolution}
    \hat{\mathcal{U}}_{\tc{i}} = \mathcal{T}_t\exp\left(-\ii\int\dd t \hat{H}_{\tc{i}}(t)\right),
\end{equation}
where $\mathcal{T}_t\exp$ denotes the time-ordering exponential with respect to the time parameter $t$.\footnote{The reason why we are emphasizing the dependence on the time coordinate used to perform the time-ordering in Eq.~\eqref{eq:intimevolution} will be made clear in Sec.~\ref{sub:covcausality}.} It is natural to take the initial state $\hat{\rho}_0$ of the system to be a product state between detector and field,
\begin{equation}
    \hat{\rho}_0 = \hat{\rho}_{\tc{f}}\otimes\hat{\rho}_{\tc{d}, 0},
\end{equation}
with $\hat{\rho}_{\tc{f}}$ and $\hat{\rho}_{\tc{d}, 0}$ being the initial states for the field and the detector, respectively. With that, the final state of the detector can be expressed as
\begin{equation}\label{eq:finalstatedetector}
    \hat{\rho}_{\tc{d}} = \Tr_{\tc{f}}\left[\hat{\mathcal{U}}_{\tc{i}}\left(\hat{\rho}_{\tc{f}}\otimes\hat{\rho}_{\tc{d}, 0}\right)\hat{\mathcal{U}}_{\tc{i}}^\dagger\right],
\end{equation}
where $\Tr_{\tc{f}}$ denotes the partial trace over the Hilbert space associated to the field.  By then judiciously engineering the interaction action~\eqref{eq:UDWCoupling} and analysing the final state of the detector~\eqref{eq:finalstatedetector}, it is possible to indirectly obtain information about features of the field. In a nutshell, this is how particle detectors can be concretely used as probes for quantum field theories.

The most commonly explored version of this otherwise fairly general setting for the interaction between localized probes and quantum fields is the so-called Unruh-DeWitt (UDW) model~\cite{Unruh1976, DeWitt}. The model consists of a localized quantum system interacting with a free, real scalar field, with the coupling to the detector taken to be linear in the field amplitude. The free action of the field is therefore
\begin{equation}\label{eq:generalKG}
    S_{\tc{f}} = -\dfrac{1}{2}\int \dd V\left(g^{ab}\nabla_a\psi\nabla_b\psi + M^2\psi^2 + \xi R\psi^2\right),
\end{equation}
and the interaction between field and detector is most commonly given by
\begin{equation}\label{eq:UDWcouplingHOprep}
    S_{\tc{i}} = \lambda \int \dd V \Lambda(\mf{x})\,\mu(\tau(\mf x))\psi(\mf x).
\end{equation}
The action~\eqref{eq:generalKG} describes a Klein-Gordon field $\psi(\mf x)$ with mass $M$ and possibly non-minimal coupling to curvature, where $R$ is the Ricci scalar of the background metric and $\xi$ is some constant. As for the detector, a number of variants for its internal degrees of freedom have been explored, with different types of probe being referred to as UDW (or UDW-like) detectors. The model first adopted by Unruh in~\cite{Unruh1976} consisted of a free particle in a box; not long after that, Unruh and Wald~\cite{Unruh-Wald} considered a slightly simpler version of the setup, where the relevant dynamics of the probe is restricted to just two of its internal energy levels. In more modern parlance, the term ``UDW detector'' without further specifications is most commonly used to describe a local probe given by a two-level system (i.e., a qubit)---see. e.g.,~\cite{JormaRigid, LoukoCurvedSpacetimes, matsasUnruh, eduardoOld, Simidzija_2020, Tjoa2020, CozzellaUDWLimit, Ivan2021}. For the purposes of this paper, however, the most relevant variant will be the harmonic-oscillator UDW detector~\cite{UnruhZurek, HuMatacs, LinHu2007, Bruschi_2013, BrownHarmonic}, which describes a probe with an internal degree of freedom $q(\tau)$ whose action is given by
\begin{equation}\label{eq:ActionHO}
    S_{\tc{d}} = \dfrac{1}{2}\int \dd\tau \left[\left(\dfrac{\dd q}{\dd\tau}\right)^2 - \omega^2 q^2\right],
\end{equation}
and we take the observable $\mu(\tau)$ that couples to the field in~\eqref{eq:UDWcouplingHOprep} to simply be $q(\tau)$, so that we have
\begin{equation}
    S_{\tc{i}} = \lambda\int\dd V\,\Lambda(\mf x) q(\tau(\mf x)) \phi(\mf x).
\end{equation}
In~\eqref{eq:ActionHO}, the parameter $\omega$ corresponds to the characteristic frequency of the harmonic oscillator, which is also the energy gap between any two consecutive energy eigenstates of the detector's free Hamiltonian in its proper frame. Note that we have absorbed a spurious mass in the Lagrangian of the harmonic oscillator in~\eqref{eq:ActionHO}---which would most commonly read $\mathcal{L}_{\tc{d}} = \frac{1}{2}(m(\dd Q/\dd \tau)^2 + m\omega^2Q^2)$---by redefining $q = \sqrt{m}Q$; this gives $q$ a normalization which matches that of a scalar field in $d+1$ dimensions, particularized to the case $d=0$ (as is well-known, quantum mechanics can equivalently be seen as a ($0+1$)-dimensional field theory). 

The main appeal of the harmonic-oscillator UDW model (as opposed, say, to the more standard qubit detector) resides in the fact that the full action of the system in this case is quadratic in all dynamical variables, and therefore the detector-field dynamics is \emph{Gaussian}. The assumption of Gaussianity vastly simplifies the description of time evolution of the detector-field system, thanks to the plethora of powerful tools and results (collectively referred to as \emph{Gaussian methods} or \emph{Gaussian quantum mechanics}) which apply to systems undergoing Gaussian dynamics~\cite{ContinuousVariablesQI, gaussianquantuminfo, Adesso2014, HacklKahler2021}. Indeed, this is one of the few known cases where the time evolution of particle detectors interacting with quantum fields can be studied nonperturbatively~\cite{Bruschi_2013, BrownHarmonic, SlowUnruh}. As we will see shortly, the harmonic-oscillator UDW detector is also the archetypical example of detector that naturally emerges from the simplest model of probe system that we can formulate based on a fully local quantum field theory. This is therefore the model of reference that we will make connection to when exploring the relation between localized quantum field theories and particle detector models later on. 
\subsection{Causality and covariance in particle detector models}\label{sub:covcausality}
In Subsection~\ref{sub:GenPD}, we made an effort to introduce the basic elements of general particle detector models in terms of action functionals and Lagrangians. This approach has a number of advantages when compared to the slightly more usual presentations in RQI, which tend to introduce detectors at the level of Hamiltonians. First, in this formulation, the framework is more immediately adapted to general relativistic settings: the setup from Eqs.~\eqref{eq:genaction}-\eqref{eq:UDWCoupling} applies unchanged to detectors in arbitrary trajectories in general spacetimes, and its generalization to the case of multiple detectors in arbitrary relative states of motion is straightforwardly obtained by simply adding the free dynamics and interaction terms of each detector at the level of the total action.\footnote{Of course, this can also be done at the Hamiltonian level; however, in that case, one has to be mindful of the fact that, in general, the Hamiltonians associated to different observers will generate time evolution with respect to distinct notions of ``time'' whenever the observers in question do not share a common frame~\cite{eduardo, us}.} In that same vein, results such as the transformation properties of the interaction Hamiltonian~\eqref{eq:interactionHamiltonian} under general changes of coordinates~\cite{eduardo, us} can be immediately derived once the interaction Hamiltonian is treated as a byproduct of the interaction action~\eqref{eq:UDWCoupling}---which is itself written in a manifestly covariant way as a spacetime integral of a scalar. Lastly, talking in terms of actions and Lagrangians also allows for a more direct connection with relativistic field theory, where describing the dynamics of interacting fields at the level of Lagrangians (as opposed to Hamiltonians) is often more natural, and generally preferred if one wishes to keep the symmetries of the system manifest. 

There is, however, an important subtlety regarding the particle detector models of the kind described in Subsection~\ref{sub:GenPD} which makes them, in a strict sense, not consistent as fully relativistic theories. The problem stems from the fact that, for generic spacetime smearings $\Lambda(\mf x)$ supported in the neighborhood of the detector's trajectory $\mf{z}(\tau)$, the coupling of a particle detector to a quantum field according to the interaction~\eqref{eq:UDWCoupling} actually violates microcausality---that is, the interaction Hamiltonian density $\hat{h}_{\tc i}(\mf x)$ prescribed by Eq.~\eqref{eq:HDensity} is such that $[\hat{h}_{\tc i}(\mf x), \hat{h}_{\tc i}(\mf y)]\neq 0$ even when the spacetime points $\mf{x}$ and $\mf{y}$ are spacelike-separated~\cite{us2}. The condition of microcausality---i.e., the property that any two local operators assigned to spacelike-separated points of spacetime must commute---guarantees that spacelike-separated events cannot influence each other, and is one of the basic requirements that a theory that respects a relativistic causal structure should satisfy.\footnote{Note that this does not preclude spacelike points from being \emph{correlated}; in fact, it is well-known that the vacuum state of any relativistic field theory is highly entangled across spatial bipartitions~\cite{vacuumEntanglement, witten}. However, entanglement alone is not enough for communication or transmission of causal signals. As the maxim goes, correlation does not imply causation.} The fact that microcausality is \emph{not} satisfied by the coupling derived from the interaction Hamiltonian~\eqref{eq:HDensity} therefore means that, strictly speaking, the UDW model generally violates relativistic causality.

The Hamiltonian density~\eqref{eq:HDensity} not commuting with itself at spacelike separation also impacts the properties of the time evolution operator~\eqref{eq:intimevolution} under general changes of coordinates. Given two Cauchy surfaces $\Sigma_1$ and $\Sigma_2$ (representing some ``initial'' and ``final'' times), there are infinitely many choices of timelike coordinate $t$ which can be used to foliate the spacetime region between $\Sigma_1$ and $\Sigma_2$ by a family of Cauchy surfaces labeled by constant values of $t$. Different choices of timelike coordinate will in general assign different temporal orders between spacelike separated events, which could in principle lead to different time evolution operators from $\Sigma_1$ to $\Sigma_2$ since Eq.~\eqref{eq:intimevolution} explicitly depends on time-ordering. Fortunately, this ambiguity in the time ordering of spacelike-separated events actually makes no difference if the theory respects microcausality: if the interaction Hamiltonian $\hat{H}_I(t)$ is the integral of a Hamiltonian density $\hat{h}_{\tc{i}}(\mf x)$ that commutes with itself at spacelike separation, the ordering assigned to spacelike-separated events is irrelevant, and any timelike coordinate $t$ used to evaluate~\eqref{eq:intimevolution} will result in the same time evolution operator. This is what happens in typical relativistic field theories, where interactions are generally local and microcausal by construction. For particle detector models with $\hat{h}_{\tc i}(\mf x)$ given by~\eqref{eq:HDensity}, however, different time-orderings will generally lead to different time evolution operators. Therefore, a purely arbitrary choice of time coordinate used to evaluate~\eqref{eq:intimevolution} can make a difference in the results obtained for the final state of the detector at $\Sigma_2$ starting from the same input state at $\Sigma_1$---and in this sense, particle detector models are not generally covariant.

These issues of the particle detector models described in~\ref{sub:GenPD} are rooted in the fact that the interaction~\eqref{eq:UDWCoupling} couples one degree of freedom of the detector (i.e., the observable $\mu_A(\tau)$) to many spacelike-separated observables of the field (i.e., 
the field observables $\mathcal{O}^A(\mf{x})$ in the support of $\Lambda(\mf{x})$ at fixed $\tau$). This creates a mild level of non-locality in the theory, since the detector at a given value of its proper time $\tau$ is able to probe points that are spacelike to its position at that time. Relatedly, the coupling~\eqref{eq:UDWCoupling} also seems to implicitly select a preferred definition of ``simultaneity''---namely, the one provided by the Fermi normal coordinate time relative to the detector's worldline $\mf{z}(\tau)$---, which in turn seems to be privileging one choice of time coordinate from the outset.\footnote{In a sense, this can also be seen as a suggestion for how to deal with the noncovariance issue in practice: the presence of a preferred observer (i.e., the detector) provides, at least for practical calculations, a preferred time coordinate relative to which the time evolution operator is most naturally computed (i.e., the time coordinate naturally associated to the detector).}

From a fundamental point of view, the mild violation of causality and breaking of covariance are certainly drawbacks of the formalism of Subsection~\ref{sub:GenPD}. However, it must be emphasized that these issues are not fatal problems for the majority of the concrete applications of particle detector models. After all, these models are not purported to be fundamental descriptions of relativistic systems all the way down to microscopic scales; rather, they are \emph{effective} descriptions of systems which, in some regimes, can be accurately approximated as a single degree of freedom smeared in a small (but finite) region of space. Particle detectors can be very useful, for instance, as tools for describing the physics of the light-matter interaction, where the role of the detector is played by atoms coupled to the electromagnetic field~\cite{richard}. In this case, the scale for the size of the detector is determined by the spread of the wavefunctions of the electron in the atom, and the approximation of assigning a single quantum degree of freedom to an atom-sized spatial region is justified as long as we cannot resolve time intervals that are of the order of (or shorter than) the time it takes for light to travel between the nucleus and the outermost electron. 

More generally, the impacts of the issues with causality and covariance on the actual predictions based on particle detector models are well-understood, and can be quantified in detail~\cite{us2, PipoFTL}.  In particular, it is possible to show that the non-locality and breaking of convariance are well under control for most cases of interest. For the typical regimes where one would expect particle detectors to be good descriptions of the phenomenology at hand (essentially, when the spatial extension of the detector in its proper frame is much smaller than all the other relevant length scales of the problem, and when we consider the dynamics for times that are longer than the light-crossing time of the detector), these effects are ultimately negligible.

From a foundational point of view, however, it would still be satisfying to have a setup for local probes interacting with quantum fields which still fully respects relativistic causality and general covariance. From the discussion in the previous paragraphs, we see that there are essentially two ways of achieving this. The first is to demand that the detector be \emph{pointlike}---i.e, the coupling between detector and field only occurs at a single point in space for each value of the detector's proper time. In other words, given the detector's trajectory $\mf{z}(\tau)$, we restrict ourselves to spacetime smearings $\Lambda(\mf x)$ of the form
\begin{equation}
    \Lambda(\mf x) = \int \dd\tau\, \chi(\tau)\dfrac{\delta^{(d+1)}(x - z(\tau))}{\sqrt{-g}},
\end{equation}
in which case the interaction~\eqref{eq:UDWCoupling} reduces to
\begin{equation}\label{eq:UDWpointlike}
    S_{\tc{i}} = \lambda\int \dd\tau\,\chi(\tau)\mu_A(\tau)\mathcal{O}^A(\mf{z}(\tau)).
\end{equation}
The function $\chi(\tau)$ modulates the time dependence of the coupling in the detector's proper frame, and is commonly referred to as the \emph{switching function}. When the coupling is of this form, at no time can the detector probe points that are spacelike to it. Therefore, none of the issues with covariance and causality arise: pointlike detectors are thus fully causal and covariant. 

Restricting to pointlike detectors can be sufficient for many purposes, especially when the size of the detector plays no particular role in the phenomenon we are interested in studying. However, the singular nature of the coupling can sometimes lead to UV divergences, and in actual physical scenarios where we may want to employ detector-based approaches, the physical system playing the role of the detector is certainly not pointlike. This leads us to the second approach for solving the causality and covariance issues, which is to assign different degrees of freedom of the probe system to different spacelike-separated points in spacetime. This essentially amounts to a description of the probe that is fundamentally based on a relativistic field theory. Explaining how to make sense of this, and showing how to systematically connect a field-theory-based description of probe systems to the more usual UDW model, will be the goal of the rest of the paper.
\section{Localized quantum fields}\label{sec:LocalQF}
In line with the philosophy of describing probes for quantum fields as fully local and relativistic systems themselves, we will start with a model for a probe formulated as a full-fledged quantum field theory. In order to be useful as a \emph{local} particle detector, we would like the field theory describing the probe to have the special property of being ``confined to a finite region of space''. The most economical way of achieving this is to take the free action of the probe field $\phi_{\tc{d}}$ (the detector) to be given by
\begin{equation}\label{eq:probeaction}
    S_{\tc{d}}[\phi_{\tc{d}}] = -\dfrac{1}{2}\int \dd V \big(g^{ab}\left(\nabla_a\phi_{\tc{d}}\right)\left(\nabla_b\phi_{\tc{d}}\right) + 2 U(\mf{x})\phi_{\tc{d}}^2\big).
\end{equation}
The first term in the action above is just the standard kinetic term for any scalar field, and the second term is an additional spacetime-dependent function $U(\mf{x})$ which acts as an external potential providing some localization profile to the probe. The intuition is that we want the potential to suppress all field configurations where $\phi_{\tc{d}}$ does not go to zero as we move away from a finite region on any spatial slice; it is in that sense that the probe field is ``confined'' to a finite region of space. 

To make this a bit more precise, we assume that the potential is such that, when restricted to any spacelike Cauchy surface $\Sigma$, $U(\mf{x})$ reaches a minimum (or a set of minima) within a region of finite volume in $\Sigma$, and $U(\mf{x})$ goes to infinity as the distance of the point $\mf{x}\in \Sigma$ to this region along $\Sigma$ goes to infinity.  Another way of describing this, as done in~\cite{QFTPD}, is to say that there is some timelike curve $\mf{z}(\tau)$ on $\mathcal{M}$ such that, for every value of the proper time parameter $\tau$, it holds that $U(\mf{x})\to +\infty$ when the squared geodesic distance between the event $\mf{x}$ and the point $\mf{z}(\tau)$, or equivalently the Synge's world function $\sigma(\mf{z}(\tau), \mf{x})$, goes to $+\infty$.\footnote{Note that Synge's world function (which is \emph{half} the squared geodesic distance between two points) is negative when evaluated between timelike-separated events, so the requirement that it grows to $+\infty$ necessarily selects only points that are spacelike to $\mf{z}(\tau)$. For more details about the formal definition, properties, and applications of Synge's world function, see~\cite{poisson}.} Given a foliation of $\mathcal{M}$ by a family of Cauchy surfaces $\Sigma_\tau$ and denoting by $R_{\Sigma_\tau}$ the smallest connected region containing all the minima of $U(\mf{x})$ in the Cauchy slice $\Sigma_\tau$, the curve $\mf{z}(\tau)$ can be roughly pictured as the trajectory of the ``center'' of the region $R_{\Sigma_\tau}$ as $\tau$ varies. 

In~\cite{QFTPD}, the function $U(\mf{x})$ is written as
\begin{equation}
    U(\mf{x}) = \dfrac{m^2}{2} + V(\mf{x}),
\end{equation}
where the constant $m^2/2$ term is interpreted as giving the mass of the field $\phi_{\tc{d}}$, and $V(\mf{x})$ models some additional external potential that confines the probe. However, since the constant mass term will play no explicit role in what follows, we will refer to $U(\mf{x})$ directly as the confining potential. This is superficially different from the nomenclature adopted in~\cite{QFTPD}, but all of the properties that we require of a confining potential clearly apply to both $U$ and $V$ interchangeably, so the slight change in convention is mostly cosmetic.

We will now make some further assumptions about the setup in order to make things simpler and more explicit. First, we will take the background spacetime to be static---i.e., we assume there is a timelike Killing vector field $\chi^a$ that is orthogonal to a family of Cauchy surfaces. With that, we can define a coordinate system $(t, \bm{x})$ on $\mathcal{M}$, where the timelike coordinate $t$ (which we will refer to as \emph{Killing time} from now on) parametrizes the flow of the Killing vector field $\chi^a$, and $\bm{x} = (x^1, \dots, x^d)$ are coordinates on the spatial slices of constant $t$. In terms of these coordinates, the timelike Killing vector takes the simple expression $\chi^a = (\partial_t)^a$, and the metric can be written as
\begin{equation}\label{eq:easymetric}
    \dd s^2 = -N^2(\bm{x})\dd t^2 + h_{ij}(\bm{x})\dd x^i \dd x^j.
\end{equation}
The function $N(\bm{x})$ is known as the \emph{lapse function}, and $h_{ij}(\bm{x})$ is the induced spatial metric on the surfaces of constant $t$. We will also assume that the confining potential $U(\mf{x})$ is invariant under the flow of the timelike Killing vector field $\chi^a$. In coordinate-independent form, this is expressed as $\chi^a\nabla_a U = 0$; in the coordinates $(t, \bm{x})$, this just means that $U$ is independent of the Killing time coordinate, $U(\mf{x}) = U(\bm{x})$. The assumption that $U$ is confining then implies that $U(\bm{x})\to+\infty$ as the distance between $\bm{x}$ and the point where the minimum of $U$ is found goes to $+\infty$.

For the metric written as in~\eqref{eq:easymetric} with the coordinates $(t, \bm{x})$, we can write
\begin{equation}
    g^{ab}\left(\nabla_a\phi_{\tc{d}}\right)\left(\nabla_b\phi_{\tc{d}}\right) = -\dfrac{1}{N^2}(\partial_t\phi_{\tc{d}})^2 + h^{ij}\left(\partial_i \phi_{\tc{d}}\right)\left(\partial_j \phi_{\tc{d}}\right),
\end{equation}
where $h^{ij}$ is the inverse of the spatial metric. Writing the volume element \mbox{$\dd V = \sqrt{-g}\,\dd^{d+1}x$} explicitly as $\dd t\,\dd^d\bm{x}\,N\sqrt{h}$, the free action of the probe then becomes
\begin{align}\label{eq:actioncoordinatesbeforeIBP}
    S_{\tc{d}}[\phi_{\tc{d}}] &= \dfrac{1}{2}\int\dd t\,\dd^d\bm{x}\dfrac{\sqrt{h}}{N}(\partial_t\phi_{\tc{d}})^2\nonumber \\
    &\,\,\,\,- \int\dd t\,\dd^d\bm{x}\,\phi_{\tc{d}}\, N\sqrt{h}\,U(\bm{x})\phi_{\tc{d}}^2\nonumber \\
    &\,\,\,-\dfrac{1}{2}\int\dd t\,\dd^d\bm{x}\,N\sqrt{h}h^{ij}\left(\partial_i \phi_{\tc{d}}\right)\left(\partial_j \phi_{\tc{d}}\right).
\end{align}
Because of the confining potential $U(\bm{x})$, in every surface at a constant value of Killing time $t$, we can restrict ourselves to field configurations $\phi_{\tc{d}}(t, \bm{x})$ that go to zero at large spacelike separation from a finite region of space. This allows us to integrate the last term of Eq.~\eqref{eq:actioncoordinatesbeforeIBP} by parts at each slice of constant $t$, throw away the boundary term, and re-write the probe action as
\begin{equation}\label{eq:actionintermediate}
    S_{\tc{d}}[\phi_{\tc{d}}] = \dfrac{1}{2}\int\dd t\,\dd^d\bm{x}\,\dfrac{\sqrt{h}}{N}\big((\partial_t\phi_{\tc{d}})^2 - \phi_{\tc{d}}E^2(\bm{x})\phi_{\tc{d}}\big),
\end{equation}
where $E^2(\bm{x})$ is a differential operator that we define as acting on field configurations $\varphi$ according to
\begin{align}\label{eq:defE2}
    E^2(\bm{x})\varphi=&\,\, 2N^2(\bm{x})U(\bm{x})\varphi - \dfrac{N}{\sqrt{h}}\partial_i\left(N\sqrt{h}h^{ij}\partial_j\varphi \right) \\
    =& \,\,N^2(\bm{x})\left( 2U(\bm{x}) - \mathcal{D}_i\mathcal{D}^i\right)\varphi  \nonumber \\
    &- \dfrac{1}{2}h^{ij}\partial_i\left(N^2\right)\partial_j\varphi. \nonumber
\end{align}
In the expression above, $\mathcal{D}_i\mathcal{D}^i$ is the spatial Laplacian on the surfaces of constant Killing time,
\begin{equation}
\mathcal{D}_i\mathcal{D}^i\varphi = \dfrac{1}{\sqrt{h}}\partial_i\left(\sqrt{h}\,h^{ij}\partial_j\varphi\right).
\end{equation}

Note that the differential operator $E^2(\bm{x})$ is independent of Killing time, since $U$, $h_{ij}$, and $N$ are. It also does not contain any time derivatives, and therefore, it is a ``purely spatial'' operator---i.e., $E^2(\bm{x})\varphi(t, \bm{x})$ evaluated on a surface $\Sigma_{t_i}$ of constant time $t=t_i$ only depends on $\varphi(t_i, \bm{x})$, and is insensitive to how $\varphi(t, \bm{x})$ varies as we move away from $\Sigma_{t_i}$. This fact makes it convenient to think of $E^2(\bm{x})$ as a differential operator only on the surfaces $\Sigma_t$ characterized by constant values of Killing time. Moreover, on each surface $\Sigma_t$, we can define an inner product on the space of scalar functions $f, g: \Sigma_t\to \mathbb{R}$ as
\begin{equation}\label{eq:innerproductSigma}
    (f, g) = \int \dd^d \bm{x}\dfrac{\sqrt{h}}{N}f(\bm{x})g(\bm{x}).
\end{equation}
The operator $E^2(\bm{x})$ is clearly symmetric with respect to the inner product~\eqref{eq:innerproductSigma}, and we will assume that there is an appropriate domain where $E^2(\bm{x})$ is self-adjoint---obtained, for instance, by only considering test functions that decay sufficiently fast at spatial infinity. Finally, we will also demand that that the space of admissible field configurations $\varphi(t, \bm{x})$ be defined such that, for any fixed Killing time $t = t_i$, the pullback of $\varphi(t, \bm{x})$ to $\Sigma_{t_i}$ is in the domain of self-adjointness of $E^2(\bm{x})$. Putting all of this together, we conclude that $E^{2}(\bm{x})$ possesses a set of eigenfunctions $\{v_{\bm{n}}(\bm{x})\}$,
\begin{equation}\label{eq:eigenvalueE2}
E^2(\bm{x})v_{\bm{n}}(\bm{x}) = \omega_{\bm{n}}^2v_{\bm{n}}(\bm{x}),
\end{equation}
which are orthonormal according to the inner product~\eqref{eq:innerproductSigma},
\begin{equation}\label{eq:orthogonality}
    (v_{\bm{n}}, v_{\bm{n}'}) =  \int \dd^d \bm{x}\dfrac{\sqrt{h}}{N}v_{\bm{n}}(\bm{x})v_{\bm{n}'}(\bm{x}) = \delta_{\bm{n}, \bm{n}'},
\end{equation}
and form a basis for the set of admissible field configurations $\varphi(t, \bm{x})$ at any fixed time $t$.\footnote{Note that in order for the probe field to be stable, the operator $E^2$ has to be positive-definite, which justifies writing its eigenvalues as the manifestly positive quantity $\omega_{\bm{n}}^2$ in Eq.~\eqref{eq:eigenvalueE2}. Evidently, this was also the reason why the operator was written in the very suggestive notation $E^2(\bm{x})$ to begin with.} The assumption that the potential $U(\bm{x})$ is confining further ensures that $E^2$ has a discrete spectrum~\cite{ReedSimon4,Simon2008}, which in turn means that the label $\bm{n}$ ranges over a discrete set of values. We are thus justified in using the orthogonality condition~\eqref{eq:orthogonality} in terms of a Kronecker delta, as opposed to a Dirac delta (which would be needed if the label $\bm{n}$ were continuous). All eigenfunctions $v_{\bm{n}}(\bm{x})$ go to zero as $U(\bm{x}) \to +\infty$, and are most strongly supported around the regions where $U(\bm{x})$ attains its minima.

Since all admissible field configurations $\phi_{\tc{d}}(t, \bm{x})$ must go to zero as we move away from the minima of $U(\bm{x})$ along any slice $\Sigma_t$ of constant Killing time $t$, we can expand $\phi_{\tc{d}}$ as
\begin{equation}\label{eq:modesum}
    \phi_{\tc{d}}(t, \bm{x}) = \sum_{{\bm n}}\phi_{\bm n}(t) v_{\bm n}(\bm{x}),
\end{equation}
where the full time dependence of the field configuration is incorporated in the amplitudes $\phi_{\bm{n}}(t)$ in the expansion above. 
Plugging this back into Eq.~\eqref{eq:actionintermediate} and using the fact that the set of functions $\{v_{\bm{n}}(\bm{x})\}$ is orthonormal in the sense of the inner product from Eq.~\eqref{eq:innerproductSigma}, the action for the probe field becomes
\begin{align}\label{eq:sumHO}
     S_{\tc{d}}[\phi_{\tc{d}}] &= \sum_{\bm{n}}S_{\bm n}[\phi_{\bm{n}}], \\
     S_{\bm n}[\phi_{\bm{n}}] &= \dfrac{1}{2}\int \dd t\, \left(\dot{\phi}_{\bm{n}}^2 - \omega^2_{\bm{n}}\phi_{\bm{n}}^2\right).\nonumber 
\end{align}
where we have defined $\dot{\phi}_{\bm{n}} \equiv \dd \phi_{\bm{n}}/\dd t$. We recognize~\eqref{eq:sumHO} as the action for a set of decoupled harmonic oscillators (which we will also refer to as \emph{modes} of the field) of frequency $\omega_{\bm{n}}$, and with amplitudes given by $\phi_{\bm{n}}(t)$, as we first wrote in~\eqref{eq:ActionHO}. 

For concreteness, assume that, on every surface of constant Killing time, the potential $U(\bm{x})$ has a unique minimum, with spatial coordinates given by $\bm{x}_0$. We can normalize the timelike Killing vector field $\chi^a$ by a constant such that $\chi^a\chi_a = -1$ at $\bm{x} = \bm{x}_0$. With this choice, the Killing time $t$ becomes a proper time parameter for the trajectory of the minimum of the potential, which, in the coordinates $(t, \bm{x})$, takes the very simple form $z^{\mu}_{\tc{d}}(t) = (t, \bm{x}_0)$. Then, we can interpret the action for the localized probe field $\phi_{\tc{d}}(\mf{x})$ as simply that of an infinite (but discrete) tower of simple harmonic oscillators with proper time $t$ and proper energy gap $\omega_{\bm{n}}$, which follow a static trajectory $\mf{z}_{\tc{d}}(t)$ in spacetime. This is ultimately the re-arrangement that will allow us to reduce the theory of a probe field coupled to a Klein-Gordon field to that of a series of particle detectors coupled to a quantum field via the UDW-like coupling that is commonly found in the literature on RQI. 

\section{From probe field to particle detectors via the path integral}\label{sec:PathIntegral}
In this section we will show how to reduce the probe model given by the localized quantum field of Sec.~\ref{sec:LocalQF} to the more standard harmonic-oscillator UDW model, using the Schwinger-Keldysh path integral.

    \subsection{Setting up the path integral}
  Consider that the probe field $\phi_{\tc{d}}$ will be used as a detector that couples to a real Klein-Gordon field $\psi$. For brevity, we will often refer to $\psi$ simply as the \emph{target} field. The joint system of probe and target field can be described by the action
  \begin{equation}
      S[\phi_{\tc{d}}, \psi] = S_{\tc{d}}[\phi_{\tc{d}}] + S_{\tc{f}}[\psi] + S_{\tc{i}}[\phi_{\tc{d}}, \psi],
  \end{equation}
  where $S_{\tc{d}}$ and $S_{\tc{f}}$ are the free actions of the $\phi_{\tc{d}}$ and $\psi$, and $S_{\tc{i}}$ encodes the interaction between the two. For the target field $\psi$, we take
  \begin{equation}\label{eq:KGfieldaction}
      S_{\tc{f}}[\psi] = -\dfrac{1}{2}\int \dd V \, \left(g^{ab}\nabla_a\psi\nabla_b\psi + M^2\psi^2 + \xi R\,\psi^2\right).
  \end{equation}
   This is the usual action of a Klein-Gordon field $\psi$ with mass $M$ and possibly non-minimal coupling to curvature that we first considered in Sec.~\ref{sub:GenPD} in Eq.~\eqref{eq:generalKG}. The probe action $S_{\tc{d}}[\phi_{\tc{d}}]$ will the exact same one that we introduced in Sec.~\ref{sec:LocalQF}, given by Eq.~\eqref{eq:probeaction}. Lastly, the interaction between the probe and target field is given by
\begin{equation}\label{eq:interaction}
    S_{\tc{i}}[\phi_{\tc{d}}, \psi] = \lambda \int \dd V\,\zeta(\mf{x})\psi(\mf{x})\phi_{\tc{d}}(\mf{x}),
\end{equation}
where $\zeta(\mf{x})$ is a function that dictates the localization of the coupling between probe and target field in spacetime, and $\lambda$ is an overall coupling constant. 

We will make use the same simplifying assumptions studied in Sec.~\ref{sec:LocalQF}---namely, we take the background metric to be static, and we assume that the confining potential of the probe is independent of the Killing time $t$ parametrizing the flow of the timelike Killing vector field $\chi^a = (\partial_t)^a$. The function $\zeta(\mf{x})$ in the interaction~\eqref{eq:interaction}, however, can still depend on $t$; this simply corresponds to the case where the strength of the interaction between the probe and the system of interest can be switched on and off with time. 

Let us now present the quantum theory of the Klein-Gordon field coupled to the probe. The central object of the theory will be taken to be the Schwinger-Keldysh path integral~\cite{SchwingerPathIntegral, KeldyshPathIntegral, FeynmanVernon}, which in our case can be written as
\begin{align}\label{eq:SchwingerKeldysh}
    \mathcal{Z}[\bm{J}, \bm{J}'] \!=& \!\!\int [\dd \Phi_f][\dd \Psi_f]\int[\dd\Phi_i][\dd\Psi_i]\int[\dd\Phi'_i][\dd\Psi'_i]\nonumber \\
    &\times \rho(\Phi_i, \Psi_i; \Phi_i', \Psi_i') \nonumber\\
    &\times Z\big[\Phi_f, \Psi_f; \Phi_i, \Psi_i|\bm{J}\big]\nonumber\\
    &\times Z\big[\Phi_f, \Psi_f; \Phi_i', \Psi_i'|\bm{J}'\big]^\ast.
\end{align}
There is quite a bit of notation to unpack in Eq.~\eqref{eq:SchwingerKeldysh}, so let us take some time to clarify what each of the terms means. 

For concreteness, consider the dynamics of the theory between two Cauchy surfaces characterized by constant values of Killing time $t_i$ and $t_f$, with $t_i<t_f$. If we want, we can take $t_i \to -\infty$ and $t_f \to +\infty$, and picture the system as evolving from the asymptotic past to the asymptotic future. The capital $\Phi$'s and $\Psi$'s in Eq.~\eqref{eq:SchwingerKeldysh} denote classical field configurations $\Phi(\bm{x})$, $\Psi(\bm{x})$ at constant-time slices for the fields $\phi_{\tc{d}}$ and $\psi$ respectively, with the configurations labeled with the subscript ``$i$'' being defined at the initial Killing time $t_i$, and those with the subscript ``$f$'' being defined at the final Killing time $t_f$. The quantity \mbox{$\rho(\Phi_i, \Psi_i; \Phi_i', \Psi_i')$} denotes matrix elements of a given initial state for the probe and target field in a basis of field eigenstates: that is, we take a state $\hat{\rho}$ describing the joint system of probe and target field at the initial Cauchy surface $\Sigma_{t_i}$, and define
\begin{equation}
    \rho(\Phi_i, \Psi_i; \Phi_i', \Psi_i') \coloneqq \bra{\Phi_i, \Psi_i}\hat{\rho}\ket{\Phi_i', \Psi_i'},
\end{equation}
where we have $\ket{\Phi_i, \Psi_i}\equiv \ket{\Phi_i}\otimes\ket{\Psi_i}$, and the kets $\ket{\Phi_i}$ and $\ket{\Psi_i}$ are states on the Hilbert spaces of probe and target field satisfying
\begin{align}
    \hat{\phi}_{\tc{d}}(t_i, \bm{x})\ket{\Phi_i} &= \Phi_i(\bm{x})\ket{\Phi_i}, \label{eq:probefieldeigenstate}\\
    \hat{\psi}(t_i, \bm{x})\ket{\Psi_i} &= \Psi_i(\bm{x})\ket{\Psi_i}.
\end{align}
In short, $\ket{\Phi_i}$ and $\ket{\Psi_i}$ are field eigenstates of $\hat{\phi}_{\tc{d}}(t_i, \bm{x})$ and $\hat{\psi}(t_i, \bm{x})$ associated to the classical field configurations $\Phi_i(\bm{x})$ and $\Psi_i(\bm{x})$ at time $t_i$.

The quantity $Z\big[\Phi_f, \Psi_f; \Phi_i, \Psi_i|\bm{J}\big]$ is the Feynman path integral, which gives transition amplitudes from initial field configurations $\Phi_i, \Psi_i$ at time $t=t_i$ to final field configurations $\Phi_f, \Psi_f$ at $t=t_f$, in the presence of sources $\bm{J} \equiv (J_{\tc{d}}, J)$ for the probe and target field, respectively. It can be expressed as
\begin{equation}\label{eq:FeynmanPathIntegral}
    Z\big[\Phi_f, \Psi_f; \Phi_i, \Psi_i\big|\bm{J}\big] = \int \mathcal{D}\phi_{\tc{d}}\mathcal{D}\psi\, e^{\ii S[\phi_{\tc{d}}, \psi| \bm{J}]}
\end{equation}
where we have abbreviated
\begin{equation}\label{eq:fullintsource}
    S[\phi_{\tc{d}}, \psi| \bm{J}] \coloneqq S[\phi_{\tc{d}}, \psi]\, + \int \!\dd V\!\left(J_{\tc{d}}(\mf{x})\phi_{\tc{d}}(\mf{x}) + J(\mf{x})\psi(\mf{x})\right)
\end{equation}
and in Eq.~\eqref{eq:FeynmanPathIntegral} the path integral is performed with the boundary conditions
\begin{align}
    \phi_{\tc{d}}(t_i, \bm{x}) &= \Phi_i(\bm{x}), \,\,\,\,\,\,\psi(t_i, \bm{x}) = \Psi_i(\bm{x}), \\
    \phi_{\tc{d}}(t_f, \bm{x}) &= \Phi_f(\bm{x}),\,\,\,\,\,\, \psi(t_f, \bm{x}) = \Psi_f(\bm{x}).
\end{align}
Finally, the several integrals over $[\dd\Phi]$'s and $[\dd\Psi]$'s in Eq.~\eqref{eq:SchwingerKeldysh} denote sums over classical field configurations (according to some measure) at the corresponding constant-time slices. 

It is important to not confuse the path integral measure $\mathcal{D}\phi_{\tc{d}}$ with the measure $[\dd\Phi]$: the former is a measure on field configurations $\phi_{\tc{d}}(t, \bm{x})$ supported on the entire spacetime region between times $t_i$ and $t_f$, whereas the latter simply ranges over fixed-time field configurations $\Phi(\bm{x})$. If we were dealing with a $(0+1)$-dimensional field theory of one scalar field $q(t)$---or equivalently, if we were doing quantum mechanics of a single bosonic particle---then $[\dd\Phi]$ would essentially reduce to a simple measure $\dd q$ on the real line, whereas $\mathcal{D}\phi_{\tc{d}}$ would become the quantum-mechanical path integral measure $\mathcal{D} q$ for a particle with one position degree of freedom $q$. We note that, from a mathematically rigorous perspective, the quantum-mechanical path integral measure is still not fully well-defined. Since this article is not devoted to solving this particular problem, we will limit ourselves to the most common physicist's treatment, where an implicit choice of path integral measure is justified \emph{a posteriori} by verifying that the results obtained via the path integral are physically reasonable. For a much less heuristic and more detailed discussion on formally defining the quantum-mechanical path integral measure, however, see~\cite{IvanPathIntegrals}.

Despite its perhaps uninviting looks, the Schwinger-Keldysh path integral~\eqref{eq:SchwingerKeldysh} is actually a rather simple object at a conceptual level. If we have the Schrodinger picture in mind\footnote{Of course, a completely analogous interpretation could be provided in terms of the Heisenberg picture, which is the one that is most useful/convenient for most purposes in field theory. I just felt the summary presented in this paragraph was more succinctly stated in the Schrodinger picture.}, we can understand Eq.~\eqref{eq:SchwingerKeldysh} as arising from the following construction (for a very detailed review, see e.g.~\cite{BenTov}). First we take some initial state $\hat{\rho}$ for the probe and target field at an early time $t_i$. Next, we define the operator $\hat{U}_{\bm{J}}(t_f, t_i)\hat{\rho}\,\hat{U}_{\bm{J}'}(t_f, t_i)^\dagger$, where $\hat{U}_{\bm{J}}(t_f, t_i)$ and $\hat{U}_{\bm{J}'}(t_f, t_i)$ are the time evolution operators from the Cauchy surface at Killing time $t_i$ to the Cauchy surface at Killing time $t_f$ in the presence of sources $\bm{J}$ and $\bm{J}'$ for the probe and target fields. We then expand the initial state on a common basis of field eigenstates for the probe and target field at time $t_i$, and take the trace of the resulting operator $\hat{U}_{\bm{J}}(t_f, t_i)\hat{\rho}\,\hat{U}_{\bm{J}'}(t_f, t_i)^\dagger$ on a basis of field eigenstates at the later time $t_f$.
The Feynman path integral appears because it corresponds precisely to the transition amplitudes between field eigenstates from early to late times---or equivalently, they are just the matrix elements of the time evolution operator in the basis of field eigenstates at fixed time slices. The addition of the sources $J_{\tc{d}}, J$ in the Feynman path integral for the probe and target just serves the purpose of allowing us to compute expectation values of observables in the evolved state, by the usual trick of taking functional derivatives with respect to the sources and then setting the sources to zero. For instance, in terms of $\mathcal{Z}[\bm{J}, \bm{J}']$, we can write
\begin{align}
    \langle\hat{\psi}(\mf{x})\rangle_{\hat{\rho}} = \dfrac{(-\ii)}{\sqrt{-g(x)}}\dfrac{\delta \mathcal{Z}}{\delta J(x)}\Bigg|_{{(\bm J}, {\bm J}') = 0}
\end{align}
\begin{align}
    \langle\hat{\psi}(\mf{x})\hat{\psi}(\mf{x}')\rangle_{\hat{\rho}} = \dfrac{1}{\sqrt{-g(x)}\sqrt{-g(x')}}\dfrac{\delta^2 \mathcal{Z}}{\delta J(x)\delta J'(x')}\Bigg|_{{(\bm J}, {\bm J}') = 0}, \label{eq:twopointfunctionSchwinger}
    \end{align}
    \begin{align}
    \big\langle\mathcal{T}\big(\hat{\psi}(\mf{x})\hat{\psi}(\mf{x}')\big)\big\rangle_{\hat{\rho}} &= \dfrac{(-\ii)^2}{\sqrt{-g(x)}\sqrt{-g(x')}}\dfrac{\delta^2 \mathcal{Z}}{\delta J(x)\delta J(x')}\Bigg|_{{(\bm J}, {\bm J}') = 0},\label{eq:timeorderedschwinger}
\end{align}
and so on. Of course, similar expressions also hold for expectation values involving the probe field, if we take functional derivatives with respect to the probe's source $J_{\tc{d}}$. We also used $\mathcal{T}$ to denote the time-ordering operation---which, based on the discussion in Subsection~\ref{sub:covcausality}, can be defined with respect to any timelike coordinate that yields the same time orientation (i.e., the same notion of past and future) as the chosen Killing time $t$, since both the probe and the target fields are fully relativistic and causal by construction~\cite{us2, PipoFTL}. It is also worth pointing out that, unlike the bare Feynman path integral that we are probably most familiar with, the Schwinger-Keldysh path integral~\eqref{eq:SchwingerKeldysh} can yield arbitrary (i.e., not necessarily time-ordered) expectation values, as emphasized, for instance, by Eq.~\eqref{eq:twopointfunctionSchwinger}. This is due to the presence of two independent sources $\bm{J}, \bm{J}'$ in the definition of~\eqref{eq:SchwingerKeldysh}---note that the only difference between~\eqref{eq:twopointfunctionSchwinger} and~\eqref{eq:timeorderedschwinger} is that the functional derivative is taken with respect to the same current $J$ in the latter case, but with respect to different currents $J$ and $J'$ in the former. For this reason, the Schwinger-Keldysh path integral is sometimes described as the \emph{expectation-value} generating functional. 

For later reference, it is sometimes also useful to introduce the so-called \emph{influence phase} $\mathcal{W}(\bm{J}, \bm{J}')$, defined such that
\begin{equation}\label{eq:influencephase}
    \mathcal{Z}[\bm{J}, \bm{J}'] = e^{\ii \mathcal{W}[\bm{J}, \bm{J}']}.
\end{equation}
Just like $\mathcal{Z}[\bm{J}, \bm{J}']$ is the generating functional for expectation values, $\mathcal{W}[\bm{J}, \bm{J}']$ is the generating functional for \emph{connected} expectation values. 

 The Schwinger-Keldysh path integral often appears with different names, depending on where it shows up in the physics literature. In many-body physics and open quantum systems, it is common to refer to it as the Feynman-Vernon influence functional~\cite{rammer_2007, calzetta_hu_2008} or decoherence functional~\cite{DecoherenceFunctional}. It also comes under the guise of the so-called \emph{closed-time-path} or \emph{in-in formalism}~\cite{CalzettaClosed, InInFormalism}, which is a general framework for the non-equilibrium dynamics of quantum systems where the Schwinger-Keldysh path integral plays a central role. These names are probably motivated by the fact that $\mathcal{Z}$ can be visualized as a path integral over two copies of the original spacetime region between the Cauchy surfaces $\Sigma_i$ and $\Sigma_f$, where the integral over the first copy ``runs forwards'' in time (form $t_i$ to $t_f$) and the integral over the second copy ``runs backwards'' (from $t_f$ to $t_i$), thus bringing us back to the initial time $t_i$ where the input state $\hat{\rho}$ was originally defined. This is in contrast with the standard Feynman path integral~\eqref{eq:FeynmanPathIntegral}, which is most readily applicable to the computation of transition amplitudes between early (in) states and late (out) states, with time only going in one direction. 

Now that we know what each of the terms in~\eqref{eq:SchwingerKeldysh} means, we can proceed with the Schwinger-Keldysh path integral for the probe-field system. We saw in Sec.~\ref{sec:LocalQF} that the action for the probe can be written as
\begin{equation}
    S_{\tc{d}}[\phi_{\tc{d}}] = \sum_{\bm{n}}\left[\dfrac{1}{2}\int \dd t \left(\dot{\phi}^2_{\bm{n}} - \omega^2_{\bm{n}}\phi_{\bm{n}}^2\right)\right],
\end{equation}
where we recall that $\phi_{\bm{n}}(t)$ are the amplitudes for the field configuration $\phi_{\tc{d}}(t, \bm{x})$ expressed as~\eqref{eq:modesum},
\begin{equation}\label{eq:modesum2}
    \phi_{\tc{d}}(t, \bm{x}) = \sum_{{\bm n}}\phi_{\bm n}(t) v_{\bm n}(\bm{x}),
\end{equation}
and $\{v_{\bm{n}}(\bm{x})\}$ is a basis of eigenfunctions of the spatial differential operator $E^2(\bm{x})$ defined in Eq.~\eqref{eq:defE2}. 
By applying the same expansion~\eqref{eq:modesum2} to the coupling of the probe with the target field $\psi$ and the probe's source $J_{\tc{d}}$ in Eq.~\eqref{eq:fullintsource}, we directly obtain
\begin{align}
    \lambda\int \dd V \,\zeta(\mf{x})\psi(\mf x)\phi_{\tc{d}}(\mf x) &= \sum_{\bm n} \lambda\int \dd t \,\phi_{\bm n}(t)\psi_{\bm n}(t), \\
    \int \dd V J_{\tc{d}}(\mf x)\phi_{\tc{d}}(\mf x) &= \sum_{\bm n} \int \dd t \,\phi_{\bm n}(t)J_{\bm n}(t)
\end{align}
where clearly
\begin{align}
    \psi_{\bm n}(t) &= \int \dd^d \bm{x}\,N\sqrt{h}\,v_{\bm n}(\bm{x})\zeta(t, \bm{x})\psi(t, \bm{x}),\\
    J_{\bm n}(t) &= \int \dd^d \bm{x}\,N\sqrt{h}\,v_{\bm n}(\bm{x})J_{\tc{d}}(t, \bm{x}).
\end{align}
 Now, from the point of view of the path integral, the expansion~\eqref{eq:modesum2} can essentially be seen as a change of dynamical variables of the probe system, from the field configurations $\phi_{\tc{d}}(\mf{x})$ to the amplitudes $\phi_{\bm n}(t)$. This helps us in two important ways. First, it allows us to schematically write the integration measure $\mathcal{D}\phi_{\tc{d}}$ in the Feynman path integral~\eqref{eq:FeynmanPathIntegral} as
\begin{equation}\label{eq:redefmeasure}
     \mathcal{D}\phi_{\tc{d}} = \mathcal{N}\prod_{\bm n} \mathcal{D}\phi_{\bm n},
\end{equation}
where $\mathcal{N}$ plays the role of the Jacobian determinant from the change of variables. It can be shown by a simple comparison argument that we must have $\mathcal{N}=1$ (see Appendix~\ref{app:computingN} for details).\footnote{One might worry that equations such as~\eqref{eq:redefmeasure} with $\mathcal{N}=1$, as well as~\eqref{eq:tensorproductmodestates} or~\eqref{eq:fixedtimemeasure}, are not dimensionally consistent---after all, the units of a scalar field in $d+1$ dimensions are certainly not those of a product of infinitely many scalar fields in $0+1$ dimensions. The reason why these equations are fine is because, as it turns out, the more rigorous definitions of both the integration measure $\mathcal{D}\phi_{\tc{d}}$ and the inner product between field configurations at a constant time (which indirectly impacts the normalizations of $\ket{\Phi}$ and $[\dd\Phi]$) actually carry implicit choices of dimensionful scales. If one carefully keeps track of these choices of scale in both $d=0$ and higher dimensions, it can be shown~\cite{IvanPathIntegrals} that things work out in such a way that~\eqref{eq:redefmeasure} (with $\mathcal{N}=1$), as well as~\eqref{eq:tensorproductmodestates} and~\eqref{eq:fixedtimemeasure}, all have the correct units. I am deeply thankful to Iv\'{a}n M. Burbano for clarifications on this.} 
 Secondly, since every field eigenstate $\ket{\Phi}$ at a surface of constant Killing time can be identified with a unique set of coefficients $\varphi_{\bm{n}}$ in the mode expansion of $\Phi(\bm{x})$ in terms of the eigenfunctions $\{v_{\bm{n}}(\bm{x})\}$, we always have
 \begin{equation}\label{eq:tensorproductmodestates}
     \ket{\Phi} = \bigotimes_{\bm{n}}\ket{\varphi_{\bm{n}}}
 \end{equation}
 where each $\ket{\varphi_{\bm{n}}}$ is an eigenstate of $\hat{\phi}_{\bm{n}}$ with eigenvalue $\varphi_{\bm{n}}$ corresponding to the component of $\Phi(\bm{x})$ in the basis $\{v_{\bm{n}}(\bm{x})\}$. 
 With this, we can write the Feynman path integral as
\begin{align}\label{eq:genfunctionalmodes}
    Z\big[\Phi_f, \Psi_f; \Phi_i, \Psi_i\big|\bm{J}\big]  &= \int \mathcal{D}\psi \,  e^{iS_{\tc{f}}}e^{i\int \dd V J(\mf{x})\psi(\mf{x})} \\
    &\times \prod_{{\bm n}} Z_{\tc{d}}[\varphi_{\bm{n}, f}; \varphi_{\bm{n}, 0}| J_{\bm{n}}  + \lambda \psi_{\bm n}] ,\nonumber
\end{align}
where we have defined
\begin{align}\label{eq:transitionamplitudemodes}
   Z_{\tc{d}}[\varphi_{\bm{n}, f}; \varphi_{\bm{n}, 0}| J_{\bm{n}}  + \lambda \psi_{\bm n}] = &\int \mathcal{D}\phi_{\bm n}\,e^{iS_{{\bm n}}[\phi_{\bm{n}}]}\\
   &\times e^{i\int \dd t \,\phi_{\bm{n}}(t)\left(\lambda\psi_{\bm{n}}(t) + J_{\bm{n}}(t)\right)}.\nonumber
\end{align}
The path integral in the target field $\psi$ is performed with the boundary conditions
\begin{equation}
    \psi(t_i, \bm{x}) = \Psi_i(\bm{x}), \,\,\,\,\psi(t_f, \bm{x}) = \Psi_f(\bm{x})
\end{equation}
and similarly, the boundary conditions for the Feynman path integrals corresponding to each mode $\phi_{\bm{n}}$ are 
\begin{equation}
    \phi_{\bm{n}}(t_i) = \varphi_{\bm{n}, 0},\,\,\,\,\phi_{\bm{n}}(t_f) = \varphi_{\bm{n}, f}
\end{equation}
where $\varphi_{\bm{n}, 0}$ and $\varphi_{\bm{n}, f}$ are the components of the probe field configurations $\Phi_i(\bm{x})$ and $\Phi_f(\bm{x})$ in terms of the basis $\{v_{\bm{n}}(\bm{x})\}$ at times $t_i$ and $t_f$, respectively. Finally, the measure $[\dd\Phi]$ on fixed-time field configurations in the Schwinger-Keldysh path integral~\eqref{eq:SchwingerKeldysh} can be simply interpreted as
\begin{equation}\label{eq:fixedtimemeasure}
    [\dd\Phi] = \prod_{\bm n}\dd\varphi_{\bm{n}}.
\end{equation}
With this, we are now finally ready to see how to reduce the probe field to a finite number of modes at the level of the Schwinger-Keldysh path integral. 
\subsection{Reduction of probe field to a finite set of modes}\label{sub:reductionfinitemodes}
Consider a subset of the probe's degrees of freedom consisting of modes indexed by $\bm{n}$ such that $\bm{n} \in A$, where $A$ comprises some finite set of labels. As a slight abuse of language and for the sake of brevity, we will often refer to these modes simply as ``the modes in $A$''. The full Hilbert space of the probe and target field (which can be described initially as $\mathcal{H} = \mathcal{H}_{\psi}\otimes\mathcal{H}_{\tc{d}}$, where $\mathcal{H}_{\psi}$ denotes the Hilbert space for the target field, and $\mathcal{H}_{\tc{d}}$ is the Hilbert space of the probe) can then be further decomposed as $\mathcal{H} = \mathcal{H}_{\psi}\otimes\mathcal{H}_{A}\otimes\mathcal{H}_{\bar{A}}$, where $\mathcal{H}_{A}$ corresponds to the subsystem comprising the modes in $A$, and $\mathcal{H}_{\bar{A}}$ contains all of the remaining modes of the probe field which are not included in $A$. 

The reduction of the theory of the probe field to a finite number of modes will consist of two steps. First, we will assume that we can only turn on sources $J_{\tc{d}}$ for the probe system such that $J_{\bm{n}} = 0$ for $\bm{n} \notin A$. Under this assumption, the Feynman path integral~\eqref{eq:genfunctionalmodes} can be written explicitly as
\begin{align}\label{eq:genfunctionalsomemodes}
    Z\big[\Phi_f, \Psi_f; \Phi_i, \Psi_i\big|\bm{J}\big] &= \int \mathcal{D}\psi \,  e^{iS_{\tc{f}}}e^{i\int \dd V J(\mf{x})\psi(\mf{x})} \\
    &\times \prod_{\bm{n}\in A} Z_{\tc{d}}[\varphi_{\bm{n}, f}; \varphi_{\bm{n}, 0}| J_{\bm{n}}  + \lambda \psi_{\bm{n}}],\nonumber \\
    &\times\prod_{{\bm n}\notin A} Z_{\tc{d}}[\varphi_{\bm{n}, f}; \varphi_{\bm{n}, 0}|\lambda \psi_{\bm n}].\nonumber
\end{align}
Next, we will assume that at early times, the field$+$probe system is initialized in a state that contains no correlations between the subsystems $\mathcal{H}_{\psi}\otimes\mathcal{H}_{A}$ and $\mathcal{H}_{\bar{A}}$. In other words, the initial state at time $t=t_i$ can be expressed as
\begin{equation}\label{eq:separableinputstate}
    \hat{\rho} = \hat{\rho}_{\psi, A}\otimes\hat{\rho}_{\bar{A}},
\end{equation}
where $\hat{\rho}_{\bar{A}}$ is a fixed state on the Hilbert space of all the modes of the probe field excluding the ones in $A$. For concreteness, we will take this state to be the vacuum---i.e., we have
\begin{equation}\label{eq:vacuumnotX}
    \hat{\rho}_{\bar{A}} = \bigotimes_{\bm{n}\notin A}\ket{0_{\bm{n}}}\bra{0_{\bm{n}}}
\end{equation}
with $\ket{0_{\bm{n}}}$ being the vacuum state of the mode $\bm{n}$. As usual, this is defined for any mode as the unique state that is annihilated by the annihilation operator associated to the mode $\bm{n}$---i.e., it is the state such that
\begin{equation}
    \hat{a}_{\bm{n}}\ket{0_{\bm n}} = 0
\end{equation}
where we define
\begin{equation}
    \hat{a}_{\bm n} = \dfrac{1}{\sqrt{2}}(\hat{\phi}_{\bm{n}} + \ii \hat{\pi}_{\bm{n}})
\end{equation}
and $\hat{\pi}_{\bm{n}} \coloneqq \dd \hat{\phi}_{\bm{n}}/\dd t$ is the canonically conjugate momentum to $\hat{\phi}_{\bm{n}}$.

If we now replace the state~\eqref{eq:separableinputstate} and the Feynman path integral~\eqref{eq:genfunctionalsomemodes} explicitly in Eq.~\eqref{eq:SchwingerKeldysh}, we find that the Schwinger-Keldysh path integral can be evaluated in the following way. For brevity of notation, we will abbreviate by $\bm{J}_{A} \equiv (\{J_{\bm{n}\in A}\}, J)$ the collection of sources for the probe$+$field system that excludes the modes not in $A$, and also use the convenient notations
\begin{align}
    \phi_{_A} &\equiv \{\phi_{\bm{n}},\,\,\bm{n}\in A\}, \\
    \dd\varphi_{_A} &\equiv \prod_{\bm{n}\in A}\dd\varphi_{\bm{n}}, \\
    \mathcal{D}\phi_{_A} &\equiv \prod_{\bm{n}\in A}\mathcal{D}\phi_{\bm{n}}, \\
    S[\phi_{A}, \psi| \bm{J}_A] &\equiv S_{\tc{f}}[\psi] + \sum_{\bm{n} \in A} S_{\bm n}[\phi_{\bm{n}}] \\
    &\,\,\,\,+ \sum_{\bm{n}\in A}\lambda \int \dd t\,\psi_{\bm{n}}(t)\phi_{\bm{n}}(t) \nonumber \\
    &\,\,\,\,+\sum_{\bm{n}\in A}\int \dd t\,J_{\bm{n}}(t)\phi_{\bm{n}}(t)+\int \dd V J(\mf{x})\psi(\mf{x}).\nonumber
\end{align}
With these conventions, we can write the Schwinger-Keldysh path integral for the modes in $A$ and the target field as
\begin{align}\label{eq:SchwingerKeldyshX}
    \mathcal{Z}[\bm{J}_{A}, \bm{J}'_A] &= \int\dd\varphi_{_A{, f}}\,\dd\varphi_{_{A, i}}\,\dd\varphi_{_{A, i}}'\nonumber \\
    &\times\int[\dd\Psi_f]\,[\dd\Psi_i]\,[\dd\Psi_i'] \nonumber \\
    &\times \rho_{\psi, A}(\varphi_{_{A, i}}, \Psi_i; \varphi_{_{A, i}}', \Psi_i')\nonumber \\
    &\times \int\mathcal{D}\phi_{_A}\mathcal{D}\phi_{_A}'\mathcal{D}\psi\,\mathcal{D}\psi' \nonumber \\
    &\times e^{\ii (S[\phi_{A}, \psi| \bm{J}_A] - S[\phi_{A}', \psi'| \bm{J}_A'])} \nonumber \\
    &\times e^{\ii \tilde{S}[\psi, \psi']},
\end{align}
where last factor above is defined as
\begin{align}\label{eq:influencephasenotX}
    e^{\ii \tilde{S}[\psi, \psi']} \equiv \prod_{\bm{n}\notin A}&\Bigg(\int\dd\varphi_{\bm{n}, f}\,\dd\varphi_{\bm{n}, i}\,\dd\varphi_{\bm{n}, i}' \nonumber \\
    &\times \braket{\varphi_{\bm{n}, i}}{0_{\bm{n}}}\braket{0_{\bm{n}}}{\varphi_{\bm{n}, i}'} \nonumber \\
    &\times \int\mathcal{D}\phi_{\bm{n}}\mathcal{D}\phi_{\bm{n}}'\nonumber \\
    &\times e^{\ii \left(S_{\bm{n}}[\phi_{\bm{n}}] + \lambda \int\dd t\, \psi_n(t)\phi_{\bm{n}}(t)\right)} \nonumber\\
    &\times e^{-\ii \left(S_{\bm{n}}[\phi_{\bm{n}}'] + \lambda \int\dd t\, \psi'_n(t)\phi'_{\bm{n}}(t)\right)}\Bigg).
\end{align}
But Eq.~\eqref{eq:influencephasenotX} is nothing but a product of Schwinger-Keldysh path integrals for the modes not included in $A$, with $\lambda\psi_{\bm{n}}$ playing the role of the source for the mode $\bm{n}$. We remember that the modes not included in $A$ are assumed to start in the ground state, which is why the factor $\braket{\varphi_{\bm{n}, i}}{0_{\bm{n}}}\braket{0_{\bm{n}}}{\varphi_{\bm{n}, i}'}$ appears in Eq.~\eqref{eq:influencephasenotX}. In this case, the quantity $\tilde{S}[\psi, \psi']$ (which is nothing but a vacuum influence phase, as we defined it in~\eqref{eq:influencephase}) can be computed exactly~\cite{BenTov}, and the result gives us 
\begin{align}\label{eq:influencephasenotX2}
    \tilde{S}[\psi, \psi'] = \dfrac{\lambda^2}{2}\sum_{\bm{n}\notin A}\int_{t_i}^{t_f}\dd t\int_{t_i}^{t_f}\dd t'&\Big(G_{\bm{n}}(t,t')\psi_{\bm n}(t)\psi_{\bm{n}}(t') \nonumber \\
    -& W_{\bm n}(t, t')\psi'_{\bm n}(t)\psi_{\bm n}(t') \nonumber \\
    -& G_{\bm n}(t, t')^\ast \psi_{\bm n}'(t)\psi_{\bm n}'(t') \nonumber \\
    +& W_{\bm n}(t, t')^\ast \psi_{\bm n}(t)\psi_{\bm n}'(t')\Big),
\end{align}
where we have defined the Wightman and Feynman functions
\begin{align}
    W_{\bm{n}}(t, t') &\coloneqq \dfrac{\ii}{2\omega_{\bm n}}e^{-\ii \omega_{\bm{n}}(t-t')}, \\
    G_{\bm{n}}(t, t')&\coloneqq \theta(t-t')W_{\bm{n}}(t, t') + \theta(t'-t)W_{\bm{n}}(t', t) \nonumber \\
    &= \dfrac{\ii}{2\omega_{\bm n}}e^{-\ii \omega_{\bm{n}}\abs{t-t'}},
\end{align}
with their complex conjugates given by
\begin{align}
    W_{\bm n}(t, t')^\ast &= -\dfrac{\ii}{2\omega_{\bm n}}e^{\ii \omega_{\bm{n}}(t-t')}, \\
    G_{\bm n}(t, t')^\ast &= -\dfrac{\ii}{2\omega_{\bm n}}e^{\ii \omega_{\bm{n}}\abs{t-t'}}.
\end{align}

Now let us step back and contemplate what we have just obtained. If it were not for the very last factor of $e^{\ii \tilde{S}[\psi, \psi']}$, the Schwinger-Keldysh path integral in Eq.~\eqref{eq:SchwingerKeldyshX} would be \emph{exactly} the same as that of a finite number of harmonic-oscillator UDW detectors with amplitudes $\phi_{\bm{n}}$ ($\bm{n} \in A$), following an orbit of the Killing vector field $\chi^a$, each coupling linearly to the target field $\psi(\mf{x})$ with the smearing functions $\Lambda_{\bm{n}}(t, \bm{x}) \coloneqq \zeta(t, \bm{x})v_{\bm{n}}(\bm x)$. This is true for any initial state $\hat{\rho}_{\psi, A}$ for the restricted system containing the modes in $A$ and the target field.
The effect of the existence of the additional modes of the probe field that have been integrated and traced out amounts to a correction that is fully encoded by the vacuum influence phase $\tilde{S}[\psi, \psi']$ for the modes not included in $A$. Note, however, that this correction is of higher order in the coupling strength $\lambda$, as made evident by Eq.~\eqref{eq:influencephasenotX2}: whereas the direct coupling between the modes in $A$ and the target field is of order $\lambda$, the net effect of the extra modes that were traced out amounts to an additional term that is of order $\lambda^2$ in the action. This means that, at lowest order in perturbation theory, if we only have access to a finite set of modes of the probe field, the physics is well-reproduced by that of a finite number of harmonic-oscillator degrees of freedom coupled to a finite set of smeared field operators. 

These results reinforce the analysis of~\cite{QFTPD}, where it was shown that, at leading order in the coupling constant and for the case where the probe field starts in its ground state, the final state of any individual mode of a full-blown field theory after all extra modes are traced out matches that of a harmonic-oscillator UDW detector. Therefore, in some sense, every statement about the dynamics of a harmonic-oscillator UDW detector can be understood as a statement about a mode of a localized quantum field. The analysis presented here vindicates and significantly extends the range of validity of this conclusion, by demonstrating that the statement remains true for an arbitrary number of modes included as detector degrees of freedom, and for a much more general set of initial states of the form~\eqref{eq:separableinputstate}. Furthermore, the full expression~\eqref{eq:SchwingerKeldyshX}, including the factor of $e^{\ii \tilde{S}[\psi, \psi']}$, also provides a way to (in principle) systematically compute the corrections that arise at higher orders in $\lambda$ from the fact that the finite set of modes fundamentally emerges from a local field theory with infinitely many degrees of freedom.

It is also clear that even the assumption that the inaccessible modes start out in the vacuum state is not essential for the conclusions above. If we had instead assumed another state $\hat{\rho}_{\bar{A}}$ for these modes, the only thing that would change is that $e^{\ii \tilde{S}[\psi, \psi']}$ in Eq.~\eqref{eq:SchwingerKeldyshX} would be the Schwinger-Keldysh path integral with sources given by $\lambda\psi_{\bm{n}}$ for the modes in $\bar{A}$, now calculated with a different input state. For a completely general $\hat{\rho}_{\bar{A}}$, this may not factorize as a product of Schwinger-Keldysh path integrals for each mode in $\bar{A}$ like it did in Eq.~\eqref{eq:influencephasenotX}; however, as long as the expectation value of the amplitudes $\hat{\phi}_{\bm n}$ for $\bm{n} \notin A$ vanish in the input state $\hat{\rho}_{\bar{A}}$, the lowest-order term in the perturbative expansion of $\tilde{S}[\psi, \psi']$ will be quadratic in the sources $\psi, \psi'$, and the lowest power of the coupling constant $\lambda$ will thus be $\lambda^2$. As such, the equivalence at leading order in perturbation theory between the dynamics of a finite number of modes of the field and the dynamics of a same number of harmonic-oscillator UDW detectors will still hold.

\section{Generalizations to multiple trajectories}\label{sec:multipleminima}
So far we have shown how to relate a description of a probe system in terms of a localized quantum field to an alternative description given by a finite number of UDW detectors following a well-defined classical trajectory in spacetime. The classical trajectory that we assign to the detectors, in turn, is directly associated to the spatial profile provided by the mode functions $v_{\bm{n}}(\bm{x})$. When $U(\bm{x})$ only has one minimum at each spatial slice of constant Killing time, it is very natural to interpret the ``trajectory of the detector'' as being simply given by the location of the minimum of the potential; after all, this will also typically be the location around which the mode functions $v_{\bm{n}}(\bm{x})$ will be mostly peaked. However, it is also possible to imagine regimes where it would be more natural to picture the potential as ``localizing the field'' in more than one spatial region---as would happen, for instance, if the confining potential actually has multiple minima which are sufficiently far apart. In this section we will briefly describe how to adapt the story from Secs.~\ref{sec:LocalQF} and~\ref{sec:PathIntegral} to this case, and comment on a few simple physical setups where this generalization can be useful.
\subsection{General strategy}\label{sub:genMultipleMinima}
For convenience, we will explicitly describe how the strategy of Section~\ref{sec:PathIntegral} can be carried over to the case where the potential generates localized profiles in two distinct regions; the generalization to more than two localized regions will be evident.

The setup is then the following. We take the potential $U(\bm{x})$ to have two distinct minima, found in two non-overlapping spatial regions $R_{\tc{a}}$ and $R_{\tc{b}}$ in each surface of constant Killing time, with $U(\bm x)\rightarrow +\infty$ as we move away from both regions. We assume that, in $R_{\tc{a}}$ and $R_{\tc{b}}$, the potential can be approximated by $U_{\tc{a}}(\bm{x})$ and $U_{\tc{b}}(\bm{x})$ respectively, where $U_{\tc{a, b}}(\bm{x})$ are both confining potentials in their own right. We can then formally write down a mode expansion for any field configuration for the probe as
\begin{equation}
    \phi_{\tc{d}}(t, \bm{x}) = \sum_{\bm{n}_{\tc{a}}}\phi^{(\tc{a})}_{\bm{n}_{\tc{a}}}(t) v^{(\tc{a})}_{\bm{n}_{\tc{a}}}(\bm{x}) + \sum_{\bm{n}_{\tc{b}}}\phi^{(\tc{b})}_{\bm{n}_{\tc{b}}}(t) v^{(\tc{b})}_{\bm{n}_{\tc{b}}}(\bm{x}),
\end{equation}
where $v^{(\tc{a, b})}_{\bm{n}_\tc{a, b}}(\bm{x})$ are mode functions corresponding to spatially localized profiles associated to the potentials $U_{\tc{a, b}}(\bm{x})$ and centered in regions $R_{\tc{a, b}}$ respectively. We will also assume that that the potential barrier and the spatial separation between the regions $R_{\tc{a}}$ and $R_{\tc{b}}$ under consideration are large enough for any two modes associated to two different regions to approximately not have any overlap; i.e., 
\begin{equation}\label{eq:nooverlapassumption}
    \int \dd^d \bm{x}\dfrac{\sqrt{h}}{N}v^{(\tc{a})}_{\bm{n}_{\tc{a}}}(\bm{x})v^{(\tc{b})}_{\bm{n}_{\tc{b}}}(\bm{x}) \simeq 0 \,\,\,\,\forall \,\,\bm{n}_{\tc{a}}, \bm{n}_{\tc{b}}.
\end{equation}
Under this assumption, the expansion of the free action for the probe field then proceeds in exactly the same way as in the previous section. The exact same steps from Sec.~\ref{sec:LocalQF} then lead to the probe action
\begin{align}\label{eq:probefieldmultipleminima}
     S_{\tc{d}} \simeq &\sum_{\bm{n}_{\tc{a}}}\dfrac{1}{2}\int \dd t\,\left[\left(\dfrac{\dd \phi^{(\tc{a})}_{\bm{n}_{\tc{a}}}}{\dd t}\right)^2 - \omega^2_{\bm{n}_{\tc{a}}}\phi^{(\tc{a})}_{\bm{n}_{\tc{a}}}{}^2\right]\\
     &+ \sum_{\bm{n}_{\tc{b}}}\dfrac{1}{2}\int \dd t\,\left[\left(\dfrac{\dd \phi^{(\tc{b})}_{\bm{n}_{\tc{b}}}}{\dd t}\right)^2 - \omega^2_{\bm{n}_{\tc{b}}}\phi^{(\tc{b})}_{\bm{n}_{\tc{b}}}{}^2\right]\nonumber.
\end{align}
In other words, the probe splits into two decoupled series of harmonic oscillators localized around two distinct regions of space. The quantization of the probe field then proceeds in a manner that is mathematically identical to what one would do in the case of two distinct fields, each under the influence of a separate confining potential.

If we do a very similar analysis of this setup using canonical quantization, we find that~\eqref{eq:nooverlapassumption} can also be interpreted as the requirement that all the quadrature operators of the modes supported in each separate region satisfy canonical commutation relations (see Appendix~\ref{app:commutationrelations}). This is consistent with the interpretation just described, in which the no-overlap assumption~\eqref{eq:nooverlapassumption} corresponds to the case where the localized quantum field can be decomposed in two independent fields that are localized around the distinct minima of the confining potential. 

As a limiting case of this picture, one could consider a situation where the regions of interest $R_{\tc{a}}$ and $R_{\tc{b}}$ act like Dirichlet cavities, with the potential being set to zero in either region, and equal to infinity everywhere else. In this limiting case, the condition~\eqref{eq:nooverlapassumption} holds exactly, since the spatial profiles are identically vanishing outside each of the cavities. If, on the other hand, the condition~\eqref{eq:nooverlapassumption} is not met, then the approximation of two independent sets of modes localized in either $R_{\tc{a}}$ or $R_{\tc{b}}$ is not good; instead, the truly independent modes will be best described as judiciously chosen linear combinations of $v_{\bm{n}_{\tc{a}}}(\bm{x})$ and $v_{\bm{n}_{\tc{b}}}(\bm{x})$ which define a set of mode functions that do satisfy the orthogonality condition~\eqref{eq:orthogonality}.

Now that we know that the free action of the probe field splits as in~\eqref{eq:probefieldmultipleminima} under the no-overlap assumption~\eqref{eq:nooverlapassumption}, we can apply exactly the same logic from Section~\ref{sec:PathIntegral}. In this case, we can have detectors at two fixed spatial positions $\bm{x}_{\tc{a}}$ and $\bm{x}_{\tc{b}}$, corresponding to the position of the minimum of the potential that each respective field mode is associated to. By tracing out all but finitely many modes in both of the regions where the field is localized, the exact same steps from Subsection~\ref{sub:reductionfinitemodes} will then give us, at leading order in perturbation theory, the same dynamics as that of a finite number of harmonic-oscillator UDW detectors supported in each individual region. If, for instance, we keep only one mode in each region (say, the lowest-frequency mode in both $R_{\tc{a}}$ and $R_{\tc{b}}$), the physics at leading order in the coupling constant is exactly the same as that of the following action,
\begin{align}\label{eq:twomodeslocalized}
    S[\phi^{(\tc{a})}_{\bm{0}}, \phi^{(\tc{b})}_{\bm{0}}, \psi] = &\,\dfrac{1}{2}\int \dd t\, \left[\left(\dfrac{\dd \phi^{(\tc{a})}_{\bm{0}}}{\dd t}\right)^2 - \omega^2_{\tc{a}, 0}\phi^{(\tc{a})}_{\bm{0}}{}^2\right] \nonumber \\
    +&\,\dfrac{1}{2}\int \dd t\, \left[\left(\dfrac{\dd \phi^{(\tc{b})}_{\bm{0}}}{\dd t}\right)^2 - \omega^2_{\tc{b}, 0}\phi^{(\tc{b})}_{\bm{0}}{}^2\right] \nonumber \\
    -& \dfrac{1}{2}\int \!\dd V \!\left(g^{ab}\nabla_a\psi\nabla_b\psi + M^2\psi^2 + \xi R\,\psi^2\right)\nonumber \\ 
    +&\, \lambda\int \dd t\left(\phi^{(\tc{a})}_{\bm{0}}(t)\psi_{\tc{a}}(t) + \phi^{(\tc{b})}_{\bm{0}}(t)\psi_{\tc{b}}(t) \right)
\end{align}
where, as before,
\begin{align}
    \psi_{\tc{a}}(t) = \int \dd^n \bm{x}\,N\sqrt{h}\,\zeta(t, \bm{x})v^{(\tc{a})}_{\bm{0}}(\bm{x})\psi(t, \bm{x}),\\
    \psi_{\tc{b}}(t) = \int \dd^n \bm{x}\,N\sqrt{h}\,\zeta(t, \bm{x})v^{(\tc{b})}_{\bm{0}}(\bm{x})\psi(t, \bm{x}).
\end{align}
With that, Eq.~\eqref{eq:twomodeslocalized} simply becomes the theory of two static harmonic-oscillator UDW detectors at the positions $\bm{x}_{\tc{a, b}},$ coupled to a quantum field with the spacetime smearing functions $\zeta(\mf{x})v^{(\tc{a, b})}_{\bm{0}}(\bm{x})$. 

A minor detail to keep in mind now is that, in general, it is not possible to normalize the Killing vector field $\chi^a$ such that $\chi^a\chi_a = -1$ in both $\bm{x}_{\tc{a}}$ and $\bm{x}_{\tc{b}}$. Instead, the best we can do in this case is to relate the proper time of each detector to the global Killing time $t$ by $\tau_{\tc{a}} = N(\bm{x}_{\tc{a}})\,t$ and $\tau_{\tc{b}} = N(\bm{x}_{\tc{b}})\,t$, where $N(\bm{x})$ is the lapse function from the general expression of the metric in static coordinates~\eqref{eq:easymetric}. Similarly, the characteristic frequencies $\Omega_{\tc{a}}$ and $\Omega_{\tc{b}}$ of each oscillator in its respective proper frame are related to the Killing frequencies $\omega_{\tc{a}, \bm{0}}$ and $\omega_{\tc{b}, \bm{0}}$ in Eq.~\eqref{eq:twomodeslocalized} by $N(\bm{x}_{\tc{a}})\Omega_{\tc{a}} = \omega_{\tc{a}, \bm{0}}$ and $N(\bm{x}_{\tc{b}})\Omega_{\tc{b}} = \omega_{\tc{b}, \bm{0}}$. These modifications, plus an additional rescaling of the field amplitude as $\sqrt{N(\bm{x_{\tc{a}}})}\,\phi^{(\tc{a})}_{\bm 0} \mapsto \phi^{(\tc{a})}_{\bm 0}$ and $\sqrt{N(\bm{x_{\tc{b}}})}\,\phi^{(\tc{b})}_{\bm 0} \mapsto \phi^{(\tc{b})}_{\bm 0}$, leave the free action of the detectors in~\eqref{eq:twomodeslocalized} in exactly the form of two harmonic-oscillator UDW detectors as written in Eq.~\eqref{eq:ActionHO}, when parametrized in terms of each detector's proper time.\footnote{It is curious, however, that this rescaling of the amplitude depending on the lapse function on each detector's position will lead to a difference in the effective coupling constants of each detector with the target field in the interaction action in the last line of Eq.~\eqref{eq:twomodeslocalized}.}
 \subsection{Examples}
We now briefly comment on simple examples where the general framework described in Subsection~\ref{sub:genMultipleMinima} may be of physical relevance.
 \subsubsection{Entanglement harvesting}

 Entanglement harvesting is a very well-known protocol in RQI by which two localized systems interacting with a quantum field can become entangled even when they only couple to the field in spacelike-separated regions~\cite{Reznik2003, Reznik1, Pozas-Kerstjens:2015}. The idea is that entanglement can be shared between the detectors even before they are able to causally communicate with each other, thanks to the pre-existing entanglement between subregions in the field theory~\cite{vacuumEntanglement, witten}.

 The most common setup for studying entanglement harvesting consists of a minimal modification of the one described in Section~\ref{sub:GenPD}. Now, instead of having just one detector, we can have two detectors (denoted from now on by $\tc{a}$ and $\tc{b}$) which are initialized in a completely uncorrelated state,
 \begin{equation}
     \hat{\rho}_{\tc{ab}, 0} = \hat{\rho}_{\tc{a}, 0}\otimes \hat{\rho}_{\tc{b}, 0},
 \end{equation}
 and are then put to evolve as in Eq.~\eqref{eq:finalstatedetector},
 \begin{equation}
     \hat{\rho}_{\tc{ab}} = \Tr_{\tc{f}}\left[\hat{\mathcal{U}}_{\tc{i}}\left(\hat{\rho}_{\tc{f}}\otimes\hat{\rho}_{\tc{ab}, 0}\right)\hat{\mathcal{U}}_{\tc{i}}^\dagger\right],
 \end{equation}
 where the time evolution operator $\hat{\mathcal{U}}_I$ now contains interaction terms corresponding to both detectors. It is usual to take $\hat{\rho}_{\tc{a}, 0}$ and $\hat{\rho}_{\tc{b}, 0}$ to be pure states (typically the ground state of each detector), as we know that initial mixedness in either of the probes generally hinders harvesting~\cite{max, RuepReply}. We then study how much entanglement the detectors have acquired, by evaluating the entanglement between the subsystems $\tc{a}$ and $\tc{b}$ in the final state $\hat{\rho}_{\tc{ab}}$. This, in turn, will usually require the use of mixed-state measures of entanglement---after all, time evolution will also generically entangle the detectors with the field, and therefore the final state $\hat{\rho}_{\tc{ab}}$ will typically be mixed.
 
 It is easy to see how the general framework of Subsection~\ref{sub:genMultipleMinima} can be directly applied to entanglement harvesting. In this case, the two parties that we are trying to entangle are both fundamentally derived from the same underlying quantum field. This mimics a setup where the systems being used as detectors consist of identical elementary particles---for instance, two electrons probing the electromagnetic field in two distinct positions. Since the free action of the probe field splits as in~\eqref{eq:probefieldmultipleminima}, the vacuum state of the probe field can be taken as a tensor product of the ground states associated to each localization region separately. In particular, if we take only the modes of lowest frequency in both localized regions considered in Subsec.~\ref{sub:genMultipleMinima}, the initial state of the probes is simply $\hat{\rho}_{\tc{ab}, 0} = \ket{0_{\tc{a}}}\bra{0_{\tc{a}}}\otimes \ket{0_{\tc{b}}}\bra{0_{\tc{b}}}$. The entanglement between the two modes after interacting with the target field can then be quantified by the usual measures adopted in entanglement harvesting---most commonly, the negativity---in the final state $\hat{\rho}_{\tc{ab}}$.

 This shows how one can conceptualize entanglement harvesting setups where each particle detector is obtained as a different localized mode of one single field. Of course, an even more direct setting for the study of entanglement harvesting using quantum fields as probes is obtained by modeling the setup with two \emph{different} localized quantum fields, each of them with one single minimum. In this case, the global Hilbert space of the joint system including both probes is guaranteed to factorize by construction, and the initial state of both probe fields on the vacuum is guaranteed to factorize between modes supported in either one of the local regions. For a more complete analysis of this problem, explicitly using modes from relativistic field theories to harvest entanglement from a quantum field, see~\cite{QFTPDHarvesting}.
 
 \subsubsection{Detector in superpositions of trajectories}
Another physically interesting scenario is achieved by simply looking at the action~\eqref{eq:twomodeslocalized} from a different point of view. Namely, we could imagine that~\eqref{eq:twomodeslocalized} describes the dynamics of the target field and \emph{one} single detector, the difference being that now the detector consists of two bosonic modes instead of one. 

If we adopt the perspective in which the two modes in $R_\tc{a}$ and $R_\tc{b}$ are taken as a single detector with two degrees of freedom, the probe is now associated to two possible trajectories, centered around the points $\bm{x}_{\tc{a}}$ and $\bm{x}_{\tc{b}}$. The interaction with the field will then promote transitions between states that are localized around either one of the trajectories, and the final state of the detector will generically contain coherences that connect one trajectory to the other. In particular, one could for instance prepare an initial state of the probe which takes the form
\begin{equation}
    \hat{\rho}_{\tc{ab}, 0} = \ket{\psi}\bra{\psi},
\end{equation}
\begin{equation}\label{eq:superposedtrajectories}
    \ket{\psi} = \dfrac{1}{\sqrt{2}}\left(\ket{1_{\tc{a}}}\otimes\ket{0_{\tc{b}}} + \ket{0_{\tc{a}}}\otimes\ket{1_{\tc{b}}} \right),
\end{equation}
where we define
\begin{equation}
    \ket{1_{\tc{a}}} = \hat{a}^\dagger_{\tc{a}}\ket{0_{\tc{a}}}, \,\, \ket{1_{\tc{b}}} = \hat{a}^\dagger_{\tc{b}}\ket{0_{\tc{b}}}
\end{equation}
and $\hat{a}_{\tc{a, b}}^\dagger$ are the creation operators for the lowest-frequency mode in regions $R_{\tc{a, b}}$.
This describes an initial superposition between a one-particle excitation that is localized around position $\bm{x}_{\tc{a}}$ and another one-particle excitation localized around position $\bm{x}_{\tc{b}}$. The final state of the probe will then involve contributions that depend on the dynamics of each degree of freedom along one single trajectory, as well as correlations that connect the degrees of freedom of the detector in both possible positions.

This shares the same spirit of other proposals found in the literature for how to treat delocalized detectors which can evolve in superpositions of trajectories~\cite{SuperTrajectories, UnruhSuperposition}; the details of the model, however, are different. In the future, it might be interesting to study how this model compares with other setups previously considered, and further investigate whether there are cases where a model for a delocalized center of mass based on localized quantum fields (along the lines of what was described here) is physically reasonable.

It is also interesting to note how the physical distinction between this example (considering superpositions of trajectories) and the previous one (which is naturally set up for entanglement harvesting) is, in some sense, subjective/context-dependent: roughly speaking, it depends on whether we postulate the existence of an external agent that can have global control over both modes in $R_\tc{a}$ and $R_\tc{b}$, or we assume that only operations that act separately on $R_\tc{a}$ or $R_\tc{b}$ are allowed. Indeed, the state~\eqref{eq:superposedtrajectories} is an entangled state between the parties at $R_\tc{a}$ and $R_\tc{b}$, so it certainly cannot be engineered by only applying local operations between the two localized regions. The framework of Sec.~\ref{sub:genMultipleMinima}, however, captures both physical scenarios at once; which of them is better applicable to a given setup at hand will be determined by other constraints of the problem being treated.

\section{Discussion}\label{sec:conclusions}
In this paper we revisited a recently proposed connection~\cite{QFTPD} between localized quantum field theories and particle detector models in RQI. More specifically, we showed how to relate the dynamics of fully relativistic probe systems (here modeled by quantum fields that are confined by some external potential) to the dynamics of a finite number of probe degrees of freedom mimicking the UDW model, with harmonic oscillators coupled to the target field. 

A series of assumptions made (namely, that the free action for the probe field was quadratic, the background metric was static, and the external potential that confined the probe was invariant under the flow of the background timelike Killing vector field) allowed us to decompose the probe field in terms of a discrete tower of decoupled modes. By tracing out all but finitely many such modes, we were able to restrict the probe field to a finite number of degrees of freedom. Doing this at the level of the Schwinger-Keldysh path integral equipped us with an analytical expression for the effective dynamics of the system containing the quantum field of interest and any finite number of modes of the probe. 

For a rather general class of initial states for the probe-field system, we showed that, at leading order in the coupling constant between probe and target field, the dynamics of any finite number of modes of the probe matches the dynamics of an equal number of harmonic-oscillator UDW detectors; in the latter case, the smearing functions for the detectors are directly related to the spatial profiles of the chosen modes of the probe field. At higher orders in the coupling, the corrections to the naive UDW model which arise from the existence of the additional modes can also be written analytically, as the Schwinger-Keldysh path integral over those modes can be computed in full detail. Therefore, the fully ``nonperturbative'' dynamics for any finite subset of modes could also be solved exactly, and the deviations from the simpler UDW model can be computed systematically to any order in perturbation theory. 

The mathematical steps performed in Subsec.~\ref{sub:reductionfinitemodes} are independent of how we choose which modes of the field should be kept as UDW detectors. For the sake of concreteness, however, a physically motivated choice would be to only include those modes whose frequencies $\omega_{\bm{n}}$ are below a certain cutoff, $\omega_{\bm n} \leq \Omega$. This gives an effective-field-theory flavor to our derivation, with the procedure performed in Subsec.~\ref{sub:reductionfinitemodes} being very close in spirit to the Wilsonian approach to the renormalization group: in this case, the frequency $\Omega$ plays the role of a UV cutoff, and the action of the finite number of harmonic-oscillator UDW detectors coupled to the field, together with the additional factor from integrating out the rest of the modes, behaves as a low-energy effective action. This is in tune with the point of view that says particle detectors should be seen as effective descriptions of systems that are fundamentally relativistic, and provides a very simple toy example where the connection between this point of view and the more technical aspects of the formalism of effective field theories can be made explicit and concrete.

This comparison to effective field theories also sheds light on some of the locality issues of particle detectors which were mentioned in Subsec.~\ref{sub:covcausality}. From the literature on Wilsonian renormalization, we know that integrating out high energy modes can introduce mild non-localities in the effective low-energy action, where the typical scale of the non-localities is related to the inverse of the cutoff in energy or momentum~\cite{polchinski1999effective, BurgessEffectiveFieldTheory}. As we discussed in Subsec.~\ref{sub:covcausality}, coupling one single degree of freedom of a detector to a smeared (non-pointlike) field operator also generally leads to violations of locality and microcausality at distance scales that are controlled by the size of the spatial smearing. Obtaining the detector model from a field theory through coarse-graining-like steps as illustrated above draws a direct connection between these two statements, and also shows how to reconcile this (only apparent) conflict between smeared particle detector models and local field theory. From this point of view, the non-locality that emerges in the detector perspective is nothing but an artifact of the coarse-graining of the probe field to a finite number of modes; by including arbitrarily many modes, of arbitrarily high frequencies, one can restore a full-fledged local field theory.

The careful reader may also have noticed that, although we started out with the target field $\psi$ being given by a free Klein-Gordon field with action~\eqref{eq:KGfieldaction} and we coupled the probe field linearly with $\psi$ through the interaction~\eqref{eq:interaction}, none of the manipulations that led to the main result in Eq.~\eqref{eq:SchwingerKeldyshX} actually depended on these two assumptions. All the steps really only depended on the free action of the \emph{probe} being quadratic. The reason why we avoided a greater level of generality is that adding self-interactions and couplings to nonlinear observables of the target field may introduce additional complications related to renormalization and quantum corrections to the action. Taking the total action $S[\phi_{\tc{d}}, \psi]$ to be quadratic in all the fundamental fields appearing in the path integral, on the other hand, allows us to not worry about these subtleties. Having said that, we find it perfectly reasonable to expect that, once these potential issues are carefully taken into account, the same logic presented here will still apply---namely, by tracing out the inaccessible degrees of freedom of the detector, the theory of a probe field coupled to any (possibly composite) operator, of any (possibly interacting) field theory, can be reduced at leading order to a theory of a finite number of particle detectors coupled to the target field.

This paper was largely motivated, on the one hand, by the important role that particle detectors play in several aspects of the interface between quantum information and QFT, and on the other hand, by a recently renewed interest in a fully local and relativistic measurement framework for QFT. In this context, we believe that the connection between particle detector models and localized quantum fields introduced in~\cite{QFTPD}, and reinforced here, is an important first step towards a more general understanding of the interplay between the more mathematically rigorous FV framework~\cite{fewster1, fewster2} and the detector-based approach to RQI. 

The present work also demonstrates the usefulness of path integral methods for the formulation of nonperturbative statements relating localized quantum field theories and particle detectors. We see this as a powerful indication of the potential insights that effective field theory tools and concepts from the renormalization group can provide to particle detector models in RQI more generally. Work in connection to this has been initiated in~\cite{Alves2023}, building on a path integral formulation for UDW detectors introduced in~\cite{Ivan2021}. It is clear, however, that we have only begun to scratch the surface on this topic. It would be very interesting to understand how to make more systematic use of effective field theory methods in the context of particle detectors, as a way of bridging the gap between effective models of local probes in RQI and more fundamental physics.

\begin{acknowledgements}
    I thank Iv\'{a}n M. Burbano for very insightful discussions, and Erickson Tjoa for thorough comments on the draft. I also acknowledge financial support from the Mike and Ophelia Lazaridis Fellowship. Research at Perimeter Institute is supported in part by the Government of Canada through the Department of Innovation, Science and Industry Canada and by the Province of Ontario through the Ministry of Colleges and Universities. 
\end{acknowledgements}
\appendix
\section{Computation of the Jacobian determinant $\mathcal{N}$ in Eq.~\eqref{eq:redefmeasure}}\label{app:computingN}
In Section~\ref{sec:PathIntegral} we argued that the expansion in Eq.~\eqref{eq:modesum2} giving the localized quantum field $\phi_{\tc{d}}$ in terms of a series of independent localized modes $\phi_{\bm{n}}$ can be seen as a change of variables in the path integral. This change of variables then induces a transformation of the measure, which can be written as
\begin{equation}
    \mathcal{D}\phi_{\tc{d}} = \mathcal{N}\prod_{\bm{n}}\mathcal{D}\phi_{\bm{n}}
\end{equation}
where, as mentioned right below Eq.~\eqref{eq:redefmeasure}, $\mathcal{N}$ is formally the Jacobian determinant of the change of variables from $\phi_{\tc{d}}(\mf{x})$ to $\phi_{\bm{n}}(t)$. 

Since the map between $\phi_{\tc{d}}(\mf{x})$ and $\phi_{\bm{n}}(t)$ is linear, $\mathcal{N}$ does not depend on any field configurations, and therefore it changes the path integral at most by some overall constant. We can thus compute $\mathcal{N}$ by comparing the results for the same path integral evaluated in two different ways, which is the strategy that we will use below.

The simplest quantity that we can compute in order to infer the factor $\mathcal{N}$ is a transition amplitude between two field eigenstates in the absence of sources,
\begin{equation}\label{eq:transitionamplitude}
    \bra{\Phi_f}\hat{U}(t_f, t_i)\ket{\Phi_i} = \int\mathcal{D}\phi_{\tc{d}}\,e^{\ii S_{\tc{d}}[\phi_{\tc{d}}]},
\end{equation}
where the path integral is performed with the boundary conditions
\begin{equation}
    \phi_{\tc{d}}(t_i, \bm{x}) = \Phi_i(\bm{x}), \,\,\,\,\,\phi_{\tc{d}}(t_f, \bm{x}) = \Phi_f(\bm{x}).
\end{equation}

If we directly perform the substitution from $\phi_{\tc{d}}(\mf{x})$ to $\phi_{\bm{n}}(t)$ as a change of variables in~\eqref{eq:transitionamplitude}, the transition amplitude becomes
\begin{equation}\label{eq:prodchangevariables}
    \bra{\Phi_f}\hat{U}(t_f, t_i)\ket{\Phi_i} =\mathcal{N}\prod_{\bm n}\left(\int\mathcal{D}\phi_{\bm n}\,e^{\ii S_{\bm{n}}[\phi_{\bm{n}}]}\right)
\end{equation}
where now the boundary conditions are
\begin{equation}\label{eq:modeBC}
    \phi_{\bm{n}}(t_i) = \varphi_{\bm{n}, i}, \,\,\,\,\, \phi_{\bm{n}}(t_f) = \varphi_{\bm{n}, f}
\end{equation}
with $\varphi_{\bm{n}, i}$ and $\varphi_{\bm{n}, f}$ being the coefficients of the expansion of the probe field configurations $\Phi_i(\bm{x})$ and $\Phi_f(\bm{x})$ in the basis $\{v_{\bm{n}}(\bm{x})\}$ at times $t_i$ and $t_f$, respectively. This is of course a consequence of the fact that every well-defined field configuration at a given constant time slice has a unique expansion in terms of the basis $\{v_{\bm n}(\bm{x})\}$.

On the other hand, the fact that every field configuration $\Phi(\bm{x})$ corresponds to a unique sequence of basis coefficients $\{\varphi_{\bm{n}}\}$ also implies that we can express the kets $\ket{\Phi_i}$ and $\ket{\Phi_f}$ as 
\begin{align}\label{eq:tensorproductconfigurations}
    \ket{\Phi_i} &= \bigotimes_{\bm{n}}\ket{\varphi_{\bm{n}, i}}, \\
    \ket{\Phi_f} &= \bigotimes_{\bm{n}}\ket{\varphi_{\bm{n}, f}},
\end{align}
and since all of the modes of the field are decoupled from each other in the action~\eqref{eq:sumHO}, the time evolution operator $\hat{U}(t_f, t_i)$ can be factored as 
\begin{equation}\label{eq:productunitaries}
    \hat{U}(t_f, t_i) = \prod_n \hat{U}_{\bm{n}}(t_f, t_i),
\end{equation}
where $\hat{U}_{\bm{n}}(t_f, t_i)$ is the time evolution operator acting on the Hilbert space of each mode $\phi_{\bm{n}}$ separately. Putting the results of Eqs.~\eqref{eq:tensorproductconfigurations}-\eqref{eq:productunitaries} together, we conclude that we can write
\begin{equation}\label{eq:prodamplitudes}
    \bra{\Phi_f}\hat{U}(t_f, t_i)\ket{\Phi_i} = \prod_{\bm{n}}\bra{\varphi_{\bm{n}, f}}\hat{U}_{\bm{n}}(t_f, t_i)\ket{\varphi_{\bm{n}, i}}.
\end{equation}
But now it is clear that each factor in the product~\eqref{eq:prodamplitudes} is again expressible as a path integral in each separate mode,
\begin{equation}\label{eq:singlemodeamplitude}
    \bra{\varphi_{\bm{n}, f}}\hat{U}_{\bm{n}}(t_f, t_i)\ket{\varphi_{\bm{n}, i}} = \int\mathcal{D}\phi_{\bm n}\,e^{\ii S_{\bm{n}}[\phi_{\bm{n}}]},
\end{equation}
with boundary conditions set again by~\eqref{eq:modeBC}. By substituting~\eqref{eq:singlemodeamplitude} back in Eq.~\eqref{eq:prodamplitudes} and comparing it with the expression~\eqref{eq:prodchangevariables} obtained by directly changing variables in the path integral, and given the fact that these transition amplitudes are not identically zero, it then follows that we must have $\mathcal{N} = 1$, as we wanted to show.

\section{Normalization~\eqref{eq:orthogonality} and canonical commutation relations}\label{app:commutationrelations}

Here we will verify that the normalization condition~\eqref{eq:orthogonality} between the mode functions $\{v_{\bm{n}}(\bm{x})\}$ ensures that $\phi_{\bm{n}}(t)$ and $\pi_{\bm{n}}(t)\coloneqq \dd \phi_{\bm{n}}/\dd t$ satisfy the canonical commutation relations of position and momentum, and therefore define a genuine mode of the field. This serves as a brief consistency check between the general steps presented in Sec.~\ref{sec:LocalQF} and the usual story of canonical quantization, and also provides useful insight to the generalization explored in Sec.~\ref{sec:multipleminima} where the probe field can be localized in multiple spatial regions.

Given the foliation of spacetime $\mathcal{M}$ by spacelike Cauchy surfaces of constant Killing time $t$, the conjugate momentum $\pi_{\tc{d}}(t, \bm{x})$ to the field $\phi_{\tc{d}}(t, \bm{x})$ can be identified directly from the action~\eqref{eq:actionintermediate} as
\begin{equation}
    \pi_{\tc{d}}(t, \bm{x}) = \dfrac{\delta S_{\tc{d}}}{\delta (\partial_t \phi_{\tc{d}})} = \dfrac{\sqrt{h}}{N}\partial_t\phi_{\tc{d}}.
\end{equation}
By expanding $\phi_{\tc{d}}$ as in~\eqref{eq:modesum} and assuming that the mode functions $\{v_{\bm{n}}(\bm{x})\}$ satisfy the orthogonality condition~\eqref{eq:orthogonality}, we can write
\begin{align}
    \phi_{\bm{n}}(t) &= \int \dd^d\bm{x} \,\dfrac{\sqrt{h}}{N}\,v_{\bm{n}}(\bm{x})\phi_{\tc{d}}(t, \bm{x}), \\
    \pi_{\bm{n}}(t) &= \int \dd^d\bm{x} \,v_{\bm{n}}(\bm{x})\pi_{\tc{d}}(t, \bm{x}).
\end{align}
Then, when $\phi_{\tc{d}}(t, \bm{x})$ and $\pi_{\tc{d}}(t, \bm{x})$ are promoted to operators $\hat{\phi}_{\tc{d}}(t, \bm{x})$ and $\hat{\pi}_{\tc{d}}(t, \bm{x})$ in the usual process of canonical quantization, the equal-time commutation relations between $\hat{\phi}_{\bm{n}}(t)$ and $\hat{\pi}_{\bm{n}}(t)$ become
\begin{align}
    [\hat{\phi}_{\bm{n}}(t), \hat{\pi}_{\bm{n}'}(t)] &= \int \dd^d\bm{x}\,\dd^d\bm{x}'\,\dfrac{\sqrt{h(\bm{x})}}{N(\bm{x})}\overbrace{[\hat{\phi}_{\tc{d}}(t, \bm{x}), \hat{\pi}_{\tc{d}}(t, \bm{x}')]}^{= \ii \delta^{(d)}(\bm{x}, \bm{x}')\hat{\mathds{1}}}\nonumber \\
    &\,\,\,\,\,\,\,\,\,\,\,\,\,\,\,\,\,\,\,\,\,\,\,\,\,\,\,\,\,\,\,\,\,\,\,\,\,\,\,\,\,\,\,\,\,\,\,\,\,\,\,\,\,\,\,\,\,\,\times v_{\bm{n}}(\bm{x})v_{\bm{n}'}(\bm{x}') \nonumber \\
    &= \ii\hat{\mathds{1}}\underbrace{\int \dd^d\bm{x}\,\dd^d\bm{x}'\,\dfrac{\sqrt{h}}{N}v_{\bm{n}}(\bm{x})v_{\bm{n}'}(\bm{x})}_{= \delta_{\bm{n}, \bm{n}'} \text{assuming~\eqref{eq:orthogonality}}} \Rightarrow\nonumber 
    \end{align}
    \begin{equation}\label{eq:ccrModeApendix}
    [\hat{\phi}_{\bm{n}}(t), \hat{\pi}_{\bm{n}'}(t)] = \ii\delta_{\bm{n}, \bm{n}'}\hat{\mathds{1}},
\end{equation}
which are nothing but the canonical commutation relations between position and momentum. This shows that the orthogonality condition~\eqref{eq:orthogonality} implies that $(\hat{\phi}_{\bm n}(t), \hat{\pi}_{\bm{n}}(t))$ defines a canonically conjugate pair characterizing a single mode of the field.

It is also possible to prove the converse statement: namely, that requiring $\hat{\phi}_{\bm{n}}(t)$ and $\hat{\pi}_{\bm{n}}(t)$ to satisfy the canonical commutation relations of position and momentum~\eqref{eq:ccrModeApendix} forces the mode functions $\{v_{\bm{n}}(\bm{x})\}$ to satisfy the normalization condition~\eqref{eq:orthogonality}. To see this, we expand $\hat{\phi}_{\tc{d}}$ and its conjugate momentum again as in~\eqref{eq:modesum},
\begin{align}
    \hat{\phi}_{\tc{d}}(t, \bm{x}) &= \sum_{\bm{n}}\hat{\phi}_{\bm{n}}(t)v_{\bm{n}}(\bm{x}), \\
    \hat{\pi}_{\tc{d}}(t, \bm{x}) &= \dfrac{\sqrt{h}}{N}\sum_{\bm{n}}\hat{\pi}_{\bm{n}}(t)v_{\bm{n}}(\bm{x}).
\end{align}
Now, using the fact that $\hat{\phi}_{\tc{d}}(t, \bm{x})$ and $\hat{\pi}_{\tc{d}}(t, \bm{x})$ must always satisfy the field's canonical commutation relations and assuming that Eq.~\eqref{eq:ccrModeApendix} holds, we get that
\begin{align}
    [\hat{\phi}_{\tc{d}}(t, \bm{x}), \hat{\pi}_{\tc{d}}(t, \bm{x}')] &=\ii  \dfrac{\sqrt{h(\bm{x'})}}{N(\bm{x'})}\sum_{\bm{n}}v_{\bm{n}}(\bm{x}) v_{\bm{n}}(\bm{x}')\hat{\mathds{1}}\nonumber \\
    &= \ii \delta^{(d)}(\bm{x} - \bm{x}')\hat{\mathds{1}},
\end{align}
which implies that
\begin{equation}\label{eq:deltaresolution}
    \dfrac{\sqrt{h(\bm{x'})}}{N(\bm{x'})}\sum_{\bm{n}}v_{\bm{n}}(\bm{x}) v_{\bm{n}}(\bm{x}') = \delta^{(d)}(\bm{x} - \bm{x}').
\end{equation}
But note that we can always express
\begin{equation}
    v_{\bm{n}}(\bm{x}) = \int\dd^d\bm{x}'\,\delta^{(d)}(\bm{x} - \bm{x}')v_{\bm{n}}(\bm{x}')
\end{equation}
which, using the fact that the Dirac delta can be written as in Eq.~\eqref{eq:deltaresolution}, becomes
\begin{equation}\label{eq:basisdecomposition}
    v_{\bm{n}}(\bm{x}) = \sum_{\bm{n}'} v_{\bm{n}'}(\bm{x}) \int \dd^d\bm{x}' \,\dfrac{\sqrt{h(\bm{x}')}}{N(\bm{x}')}v_{\bm{n}'}(\bm{x}')v_{\bm{n}}(\bm{x}').
\end{equation}
Now we note that Eq.~\eqref{eq:basisdecomposition} provides an expression for the mode function $v_{\bm{n}}(\bm{x})$ as a linear combination of the other mode functions $v_{\bm{n}'}(\bm{x})$. But if $\{v_{\bm{n}}(\bm{x})\}$ forms a basis, the only possible linear combination that expresses $v_{\bm{n}}(\bm{x})$ in terms of the set $\{v_{\bm{n}'}(\bm{x})\}$ is the trivial one, where the expansion coefficients are equal to $1$ when $\bm{n} = \bm{n}'$, and zero otherwise. In short,
\begin{equation}
    \int \dd^d\bm{x}' \,\dfrac{\sqrt{h(\bm{x}')}}{N(\bm{x}')}v_{\bm{n}'}(\bm{x}')v_{\bm{n}}(\bm{x}') = \delta_{\bm{n}', \bm{n}}
\end{equation}
which is precisely~\eqref{eq:orthogonality}.

This explicitly shows that the set of modes constructed in Sec.~\ref{sec:LocalQF} matches precisely the set of normal modes that we would have constructed in the process of canonical quantization of the field $\phi_{\tc{d}}(\mf x)$. Conversely, it also shows that if the no-overlap condition~\eqref{eq:nooverlapassumption} is not satisfied, the amplitudes associated to each mode function $v_{\bm n}(\bm x)$ cannot be interpreted as independent harmonic oscillators. One can interpret this as the canonical-quantization version of the statement alluded to in Subsec.~\ref{sub:genMultipleMinima}, in which the approximation~\eqref{eq:nooverlapassumption} was crucial for justifying the statement that a localized quantum field theory where the confining potential had multiple minima could be seen as composed of two sets of independent modes supported around each of these minima.
\twocolumngrid
\bibliography{references.bib}

\begin{thebibliography}{73}%
\makeatletter
\providecommand \@ifxundefined [1]{%
 \@ifx{#1\undefined}
}%
\providecommand \@ifnum [1]{%
 \ifnum #1\expandafter \@firstoftwo
 \else \expandafter \@secondoftwo
 \fi
}%
\providecommand \@ifx [1]{%
 \ifx #1\expandafter \@firstoftwo
 \else \expandafter \@secondoftwo
 \fi
}%
\providecommand \natexlab [1]{#1}%
\providecommand \enquote  [1]{``#1''}%
\providecommand \bibnamefont  [1]{#1}%
\providecommand \bibfnamefont [1]{#1}%
\providecommand \citenamefont [1]{#1}%
\providecommand \href@noop [0]{\@secondoftwo}%
\providecommand \href [0]{\begingroup \@sanitize@url \@href}%
\providecommand \@href[1]{\@@startlink{#1}\@@href}%
\providecommand \@@href[1]{\endgroup#1\@@endlink}%
\providecommand \@sanitize@url [0]{\catcode `\\12\catcode `\$12\catcode `\&12\catcode `\#12\catcode `\^12\catcode `\_12\catcode `\%12\relax}%
\providecommand \@@startlink[1]{}%
\providecommand \@@endlink[0]{}%
\providecommand \url  [0]{\begingroup\@sanitize@url \@url }%
\providecommand \@url [1]{\endgroup\@href {#1}{\urlprefix }}%
\providecommand \urlprefix  [0]{URL }%
\providecommand \Eprint [0]{\href }%
\providecommand \doibase [0]{https://doi.org/}%
\providecommand \selectlanguage [0]{\@gobble}%
\providecommand \bibinfo  [0]{\@secondoftwo}%
\providecommand \bibfield  [0]{\@secondoftwo}%
\providecommand \translation [1]{[#1]}%
\providecommand \BibitemOpen [0]{}%
\providecommand \bibitemStop [0]{}%
\providecommand \bibitemNoStop [0]{.\EOS\space}%
\providecommand \EOS [0]{\spacefactor3000\relax}%
\providecommand \BibitemShut  [1]{\csname bibitem#1\endcsname}%
\let\auto@bib@innerbib\@empty
\bibitem [{\citenamefont {Perche}\ \emph {et~al.}(2024{\natexlab{a}})\citenamefont {Perche}, \citenamefont {Polo-G\'omez}, \citenamefont {Torres},\ and\ \citenamefont {Mart\'{\i}n-Mart\'{\i}nez}}]{QFTPD}%
  \BibitemOpen
  \bibfield  {author} {\bibinfo {author} {\bibfnamefont {T.~R.}\ \bibnamefont {Perche}}, \bibinfo {author} {\bibfnamefont {J.}~\bibnamefont {Polo-G\'omez}}, \bibinfo {author} {\bibfnamefont {B.~d. S.~L.}\ \bibnamefont {Torres}},\ and\ \bibinfo {author} {\bibfnamefont {E.}~\bibnamefont {Mart\'{\i}n-Mart\'{\i}nez}},\ }\bibfield  {title} {\bibinfo {title} {Particle detectors from localized quantum field theories},\ }\href {https://doi.org/10.1103/PhysRevD.109.045013} {\bibfield  {journal} {\bibinfo  {journal} {Phys. Rev. D}\ }\textbf {\bibinfo {volume} {109}},\ \bibinfo {pages} {045013} (\bibinfo {year} {2024}{\natexlab{a}})}\BibitemShut {NoStop}%
\bibitem [{\citenamefont {Polo-G\'omez}\ \emph {et~al.}(2022)\citenamefont {Polo-G\'omez}, \citenamefont {Garay},\ and\ \citenamefont {Mart\'{\i}n-Mart\'{\i}nez}}]{jose}%
  \BibitemOpen
  \bibfield  {author} {\bibinfo {author} {\bibfnamefont {J.}~\bibnamefont {Polo-G\'omez}}, \bibinfo {author} {\bibfnamefont {L.~J.}\ \bibnamefont {Garay}},\ and\ \bibinfo {author} {\bibfnamefont {E.}~\bibnamefont {Mart\'{\i}n-Mart\'{\i}nez}},\ }\bibfield  {title} {\bibinfo {title} {A detector-based measurement theory for quantum field theory},\ }\href {https://doi.org/10.1103/PhysRevD.105.065003} {\bibfield  {journal} {\bibinfo  {journal} {Phys. Rev. D}\ }\textbf {\bibinfo {volume} {105}},\ \bibinfo {pages} {065003} (\bibinfo {year} {2022})}\BibitemShut {NoStop}%
\bibitem [{\citenamefont {Reznik}(2003)}]{Reznik2003}%
  \BibitemOpen
  \bibfield  {author} {\bibinfo {author} {\bibfnamefont {B.}~\bibnamefont {Reznik}},\ }\bibfield  {title} {{\selectlanguage {English}\bibinfo {title} {Entanglement from the vacuum}},\ }\href {https://doi.org/10.1023/A:1022875910744} {\bibfield  {journal} {\bibinfo  {journal} {Found. Phys.}\ }\textbf {\bibinfo {volume} {33}},\ \bibinfo {pages} {167} (\bibinfo {year} {2003})}\BibitemShut {NoStop}%
\bibitem [{\citenamefont {Reznik}\ \emph {et~al.}(2005)\citenamefont {Reznik}, \citenamefont {Retzker},\ and\ \citenamefont {Silman}}]{Reznik1}%
  \BibitemOpen
  \bibfield  {author} {\bibinfo {author} {\bibfnamefont {B.}~\bibnamefont {Reznik}}, \bibinfo {author} {\bibfnamefont {A.}~\bibnamefont {Retzker}},\ and\ \bibinfo {author} {\bibfnamefont {J.}~\bibnamefont {Silman}},\ }\bibfield  {title} {\bibinfo {title} {Violating {B}ell's inequalities in vacuum},\ }\href {http://link.aps.org/abstract/PRA/v71/e042104} {\bibfield  {journal} {\bibinfo  {journal} {Phys. Rev. A}\ }\textbf {\bibinfo {volume} {71}},\ \bibinfo {eid} {042104} (\bibinfo {year} {2005})}\BibitemShut {NoStop}%
\bibitem [{\citenamefont {Pozas-Kerstjens}\ and\ \citenamefont {Mart\'{i}n-Mart\'{i}nez}(2015)}]{Pozas-Kerstjens:2015}%
  \BibitemOpen
  \bibfield  {author} {\bibinfo {author} {\bibfnamefont {A.}~\bibnamefont {Pozas-Kerstjens}}\ and\ \bibinfo {author} {\bibfnamefont {E.}~\bibnamefont {Mart\'{i}n-Mart\'{i}nez}},\ }\bibfield  {title} {\bibinfo {title} {Harvesting correlations from the quantum vacuum},\ }\href {https://doi.org/10.1103/PhysRevD.92.064042} {\bibfield  {journal} {\bibinfo  {journal} {Phys. Rev. D}\ }\textbf {\bibinfo {volume} {92}},\ \bibinfo {pages} {064042} (\bibinfo {year} {2015})}\BibitemShut {NoStop}%
\bibitem [{\citenamefont {de~S.~L.~Torres}\ \emph {et~al.}(2023)\citenamefont {de~S.~L.~Torres}, \citenamefont {Wurtz}, \citenamefont {Polo-G{\'o}mez},\ and\ \citenamefont {Mart{\'i}n-Mart{\'i}nez}}]{kelly}%
  \BibitemOpen
  \bibfield  {author} {\bibinfo {author} {\bibfnamefont {B.}~\bibnamefont {de~S.~L.~Torres}}, \bibinfo {author} {\bibfnamefont {K.}~\bibnamefont {Wurtz}}, \bibinfo {author} {\bibfnamefont {J.}~\bibnamefont {Polo-G{\'o}mez}},\ and\ \bibinfo {author} {\bibfnamefont {E.}~\bibnamefont {Mart{\'i}n-Mart{\'i}nez}},\ }\bibfield  {title} {\bibinfo {title} {Entanglement structure of quantum fields through local probes},\ }\href {https://doi.org/10.1007/JHEP05(2023)058} {\bibfield  {journal} {\bibinfo  {journal} {J. High Energy Phys.}\ }\textbf {\bibinfo {volume} {2023}}\bibinfo  {number} { (5)},\ \bibinfo {pages} {58}}\BibitemShut {NoStop}%
\bibitem [{\citenamefont {Unruh}(1976)}]{Unruh1976}%
  \BibitemOpen
\bibfield  {number} {  }\bibfield  {author} {\bibinfo {author} {\bibfnamefont {W.~G.}\ \bibnamefont {Unruh}},\ }\bibfield  {title} {\bibinfo {title} {Notes on black-hole evaporation},\ }\href {https://doi.org/10.1103/PhysRevD.14.870} {\bibfield  {journal} {\bibinfo  {journal} {Phys. Rev. D}\ }\textbf {\bibinfo {volume} {14}},\ \bibinfo {pages} {870} (\bibinfo {year} {1976})}\BibitemShut {NoStop}%
\bibitem [{\citenamefont {Candelas}\ and\ \citenamefont {Sciama}(1977)}]{Sciama1977}%
  \BibitemOpen
  \bibfield  {author} {\bibinfo {author} {\bibfnamefont {P.}~\bibnamefont {Candelas}}\ and\ \bibinfo {author} {\bibfnamefont {D.~W.}\ \bibnamefont {Sciama}},\ }\bibfield  {title} {\bibinfo {title} {Irreversible thermodynamics of black holes},\ }\href {https://doi.org/10.1103/PhysRevLett.38.1372} {\bibfield  {journal} {\bibinfo  {journal} {Phys. Rev. Lett.}\ }\textbf {\bibinfo {volume} {38}},\ \bibinfo {pages} {1372} (\bibinfo {year} {1977})}\BibitemShut {NoStop}%
\bibitem [{\citenamefont {DeWitt}(1980)}]{DeWitt}%
  \BibitemOpen
  \bibfield  {author} {\bibinfo {author} {\bibfnamefont {B.}~\bibnamefont {DeWitt}},\ }\href@noop {} {\emph {\bibinfo {title} {General Relativity; an Einstein Centenary Survey}}}\ (\bibinfo  {publisher} {Cambridge University Press},\ \bibinfo {address} {Cambridge, UK},\ \bibinfo {year} {1980})\BibitemShut {NoStop}%
\bibitem [{\citenamefont {Lopp}\ and\ \citenamefont {Mart\'{i}n-Mart\'{i}nez}(2021)}]{richard}%
  \BibitemOpen
  \bibfield  {author} {\bibinfo {author} {\bibfnamefont {R.}~\bibnamefont {Lopp}}\ and\ \bibinfo {author} {\bibfnamefont {E.}~\bibnamefont {Mart\'{i}n-Mart\'{i}nez}},\ }\bibfield  {title} {\bibinfo {title} {Quantum delocalization, gauge, and quantum optics: Light-matter interaction in relativistic quantum information},\ }\href {https://doi.org/10.1103/PhysRevA.103.013703} {\bibfield  {journal} {\bibinfo  {journal} {Phys. Rev. A}\ }\textbf {\bibinfo {volume} {103}},\ \bibinfo {pages} {013703} (\bibinfo {year} {2021})}\BibitemShut {NoStop}%
\bibitem [{\citenamefont {Sorkin}(1993)}]{Sorkin}%
  \BibitemOpen
  \bibfield  {author} {\bibinfo {author} {\bibfnamefont {R.~D.}\ \bibnamefont {Sorkin}},\ }\href@noop {} {\bibinfo {title} {Impossible measurements on quantum fields}} (\bibinfo {year} {1993}),\ \Eprint {https://arxiv.org/abs/gr-qc/9302018} {arXiv:gr-qc/9302018 [gr-qc]} \BibitemShut {NoStop}%
\bibitem [{\citenamefont {Dowker}(2011)}]{dowker2011}%
  \BibitemOpen
  \bibfield  {author} {\bibinfo {author} {\bibfnamefont {F.}~\bibnamefont {Dowker}},\ }\href@noop {} {\bibinfo {title} {Useless qubits in ``relativistic quantum information"}} (\bibinfo {year} {2011}),\ \Eprint {https://arxiv.org/abs/1111.2308} {arXiv:1111.2308 [quant-ph]} \BibitemShut {NoStop}%
\bibitem [{\citenamefont {Borsten}\ \emph {et~al.}(2021)\citenamefont {Borsten}, \citenamefont {Jubb},\ and\ \citenamefont {Kells}}]{impossibleRevisited}%
  \BibitemOpen
  \bibfield  {author} {\bibinfo {author} {\bibfnamefont {L.}~\bibnamefont {Borsten}}, \bibinfo {author} {\bibfnamefont {I.}~\bibnamefont {Jubb}},\ and\ \bibinfo {author} {\bibfnamefont {G.}~\bibnamefont {Kells}},\ }\bibfield  {title} {\bibinfo {title} {Impossible measurements revisited},\ }\href {https://doi.org/10.1103/PhysRevD.104.025012} {\bibfield  {journal} {\bibinfo  {journal} {Phys. Rev. D}\ }\textbf {\bibinfo {volume} {104}},\ \bibinfo {pages} {025012} (\bibinfo {year} {2021})}\BibitemShut {NoStop}%
\bibitem [{\citenamefont {Fewster}\ and\ \citenamefont {Verch}(2020)}]{fewster1}%
  \BibitemOpen
  \bibfield  {author} {\bibinfo {author} {\bibfnamefont {C.~J.}\ \bibnamefont {Fewster}}\ and\ \bibinfo {author} {\bibfnamefont {R.}~\bibnamefont {Verch}},\ }\bibfield  {title} {\bibinfo {title} {Quantum fields and local measurements},\ }\href {https://doi.org/10.1007/s00220-020-03800-6} {\bibfield  {journal} {\bibinfo  {journal} {Commun. Math. Phys.}\ }\textbf {\bibinfo {volume} {378}},\ \bibinfo {pages} {851} (\bibinfo {year} {2020})}\BibitemShut {NoStop}%
\bibitem [{\citenamefont {Fewster}\ and\ \citenamefont {Verch}(2023)}]{fewster2}%
  \BibitemOpen
  \bibfield  {author} {\bibinfo {author} {\bibfnamefont {C.~J.}\ \bibnamefont {Fewster}}\ and\ \bibinfo {author} {\bibfnamefont {R.}~\bibnamefont {Verch}},\ }\href@noop {} {\bibinfo {title} {Measurement in quantum field theory}} (\bibinfo {year} {2023}),\ \Eprint {https://arxiv.org/abs/2304.13356} {arXiv:2304.13356 [math-ph]} \BibitemShut {NoStop}%
\bibitem [{\citenamefont {Bostelmann}\ \emph {et~al.}(2021)\citenamefont {Bostelmann}, \citenamefont {Fewster},\ and\ \citenamefont {Ruep}}]{impossibleImpossible}%
  \BibitemOpen
  \bibfield  {author} {\bibinfo {author} {\bibfnamefont {H.}~\bibnamefont {Bostelmann}}, \bibinfo {author} {\bibfnamefont {C.~J.}\ \bibnamefont {Fewster}},\ and\ \bibinfo {author} {\bibfnamefont {M.~H.}\ \bibnamefont {Ruep}},\ }\bibfield  {title} {\bibinfo {title} {Impossible measurements require impossible apparatus},\ }\href {https://doi.org/10.1103/PhysRevD.103.025017} {\bibfield  {journal} {\bibinfo  {journal} {Phys. Rev. D}\ }\textbf {\bibinfo {volume} {103}},\ \bibinfo {pages} {025017} (\bibinfo {year} {2021})}\BibitemShut {NoStop}%
\bibitem [{\citenamefont {Fewster}\ \emph {et~al.}(2023)\citenamefont {Fewster}, \citenamefont {Jubb},\ and\ \citenamefont {Ruep}}]{singularOrNonlocal}%
  \BibitemOpen
  \bibfield  {author} {\bibinfo {author} {\bibfnamefont {C.~J.}\ \bibnamefont {Fewster}}, \bibinfo {author} {\bibfnamefont {I.}~\bibnamefont {Jubb}},\ and\ \bibinfo {author} {\bibfnamefont {M.~H.}\ \bibnamefont {Ruep}},\ }\bibfield  {title} {\bibinfo {title} {Asymptotic measurement schemes for every observable of a quantum field theory},\ }\href {https://doi.org/10.1007/s00023-022-01239-0} {\bibfield  {journal} {\bibinfo  {journal} {Ann. Henri Poincar\'{e}}\ }\textbf {\bibinfo {volume} {24}},\ \bibinfo {pages} {1137} (\bibinfo {year} {2023})}\BibitemShut {NoStop}%
\bibitem [{\citenamefont {de~Ram\'on}\ \emph {et~al.}(2021)\citenamefont {de~Ram\'on}, \citenamefont {Papageorgiou},\ and\ \citenamefont {Mart\'{\i}n-Mart\'{\i}nez}}]{PipoFTL}%
  \BibitemOpen
  \bibfield  {author} {\bibinfo {author} {\bibfnamefont {J.}~\bibnamefont {de~Ram\'on}}, \bibinfo {author} {\bibfnamefont {M.}~\bibnamefont {Papageorgiou}},\ and\ \bibinfo {author} {\bibfnamefont {E.}~\bibnamefont {Mart\'{\i}n-Mart\'{\i}nez}},\ }\bibfield  {title} {\bibinfo {title} {Relativistic causality in particle detector models: Faster-than-light signaling and impossible measurements},\ }\href {https://doi.org/10.1103/PhysRevD.103.085002} {\bibfield  {journal} {\bibinfo  {journal} {Phys. Rev. D}\ }\textbf {\bibinfo {volume} {103}},\ \bibinfo {pages} {085002} (\bibinfo {year} {2021})}\BibitemShut {NoStop}%
\bibitem [{\citenamefont {Grimmer}(2023)}]{Dan2023}%
  \BibitemOpen
  \bibfield  {author} {\bibinfo {author} {\bibfnamefont {D.}~\bibnamefont {Grimmer}},\ }\bibfield  {title} {\bibinfo {title} {The pragmatic {QFT} measurement problem and the need for a {H}eisenberg-like cut in {QFT}},\ }\href {https://doi.org/10.1007/s11229-023-04301-4} {\bibfield  {journal} {\bibinfo  {journal} {Synthese}\ }\textbf {\bibinfo {volume} {202}},\ \bibinfo {pages} {104} (\bibinfo {year} {2023})}\BibitemShut {NoStop}%
\bibitem [{\citenamefont {Papageorgiou}\ and\ \citenamefont {Fraser}(2023)}]{MariaDoreen2023}%
  \BibitemOpen
  \bibfield  {author} {\bibinfo {author} {\bibfnamefont {M.}~\bibnamefont {Papageorgiou}}\ and\ \bibinfo {author} {\bibfnamefont {D.}~\bibnamefont {Fraser}},\ }\href@noop {} {\bibinfo {title} {Eliminating the ``impossible": Recent progress on local measurement theory for quantum field theory}} (\bibinfo {year} {2023}),\ \Eprint {https://arxiv.org/abs/2307.08524} {arXiv:2307.08524 [quant-ph]} \BibitemShut {NoStop}%
\bibitem [{\citenamefont {Schwinger}(1961)}]{SchwingerPathIntegral}%
  \BibitemOpen
  \bibfield  {author} {\bibinfo {author} {\bibfnamefont {J.}~\bibnamefont {Schwinger}},\ }\bibfield  {title} {\bibinfo {title} {{Brownian Motion of a Quantum Oscillator}},\ }\href {https://doi.org/10.1063/1.1703727} {\bibfield  {journal} {\bibinfo  {journal} {J. Math. Phys.}\ }\textbf {\bibinfo {volume} {2}},\ \bibinfo {pages} {407} (\bibinfo {year} {1961})}\BibitemShut {NoStop}%
\bibitem [{\citenamefont {Keldysh}(1965)}]{KeldyshPathIntegral}%
  \BibitemOpen
  \bibfield  {author} {\bibinfo {author} {\bibfnamefont {L.~V.}\ \bibnamefont {Keldysh}},\ }\bibfield  {title} {\bibinfo {title} {{Diagram Technique for Nonequilibrium Processes}},\ }\href {http://www.jetp.ras.ru/cgi-bin/e/index/e/20/4/p1018?a=list} {\bibfield  {journal} {\bibinfo  {journal} {Zh. Eksp. Teor. Fiz.}\ }\textbf {\bibinfo {volume} {47}},\ \bibinfo {pages} {1515} (\bibinfo {year} {1965})}\BibitemShut {NoStop}%
\bibitem [{\citenamefont {Feynman}\ and\ \citenamefont {Vernon}(1963)}]{FeynmanVernon}%
  \BibitemOpen
  \bibfield  {author} {\bibinfo {author} {\bibfnamefont {R.}~\bibnamefont {Feynman}}\ and\ \bibinfo {author} {\bibfnamefont {F.}~\bibnamefont {Vernon}},\ }\bibfield  {title} {\bibinfo {title} {The theory of a general quantum system interacting with a linear dissipative system},\ }\href {https://doi.org/https://doi.org/10.1016/0003-4916(63)90068-X} {\bibfield  {journal} {\bibinfo  {journal} {Ann. Phys.}\ }\textbf {\bibinfo {volume} {24}},\ \bibinfo {pages} {118} (\bibinfo {year} {1963})}\BibitemShut {NoStop}%
\bibitem [{\citenamefont {Anastopoulos}\ \emph {et~al.}(2023{\natexlab{a}})\citenamefont {Anastopoulos}, \citenamefont {Hu},\ and\ \citenamefont {Savvidou}}]{charis2023}%
  \BibitemOpen
  \bibfield  {author} {\bibinfo {author} {\bibfnamefont {C.}~\bibnamefont {Anastopoulos}}, \bibinfo {author} {\bibfnamefont {B.-L.}\ \bibnamefont {Hu}},\ and\ \bibinfo {author} {\bibfnamefont {K.}~\bibnamefont {Savvidou}},\ }\bibfield  {title} {\bibinfo {title} {Towards a field-theory based relativistic quantum information},\ }\href {https://doi.org/10.1088/1742-6596/2533/1/012004} {\bibfield  {journal} {\bibinfo  {journal} {J. Phys. Conf. Ser.}\ }\textbf {\bibinfo {volume} {2533}},\ \bibinfo {pages} {012004} (\bibinfo {year} {2023}{\natexlab{a}})}\BibitemShut {NoStop}%
\bibitem [{\citenamefont {Anastopoulos}\ \emph {et~al.}(2023{\natexlab{b}})\citenamefont {Anastopoulos}, \citenamefont {Hu},\ and\ \citenamefont {Savvidou}}]{charis2023Again}%
  \BibitemOpen
  \bibfield  {author} {\bibinfo {author} {\bibfnamefont {C.}~\bibnamefont {Anastopoulos}}, \bibinfo {author} {\bibfnamefont {B.-L.}\ \bibnamefont {Hu}},\ and\ \bibinfo {author} {\bibfnamefont {K.}~\bibnamefont {Savvidou}},\ }\bibfield  {title} {\bibinfo {title} {Quantum field theory based quantum information: Measurements and correlations},\ }\href {https://doi.org/https://doi.org/10.1016/j.aop.2023.169239} {\bibfield  {journal} {\bibinfo  {journal} {Ann. Phys.}\ }\textbf {\bibinfo {volume} {450}},\ \bibinfo {pages} {169239} (\bibinfo {year} {2023}{\natexlab{b}})}\BibitemShut {NoStop}%
\bibitem [{\citenamefont {Polchinski}(1999)}]{polchinski1999effective}%
  \BibitemOpen
  \bibfield  {author} {\bibinfo {author} {\bibfnamefont {J.}~\bibnamefont {Polchinski}},\ }\href@noop {} {\bibinfo {title} {{Effective Field Theory and the Fermi Surface}}} (\bibinfo {year} {1999}),\ \Eprint {https://arxiv.org/abs/hep-th/9210046} {arXiv:hep-th/9210046 [hep-th]} \BibitemShut {NoStop}%
\bibitem [{\citenamefont {Burgess}(2020)}]{BurgessEffectiveFieldTheory}%
  \BibitemOpen
  \bibfield  {author} {\bibinfo {author} {\bibfnamefont {C.~P.}\ \bibnamefont {Burgess}},\ }\href {https://doi.org/https://doi.org/10.1017/9781139048040} {\emph {\bibinfo {title} {Introduction to Effective Field Theory: Thinking Effectively about Hierarchies of Scale}}}\ (\bibinfo  {publisher} {Cambridge University Press},\ \bibinfo {year} {2020})\BibitemShut {NoStop}%
\bibitem [{\citenamefont {Foo}\ \emph {et~al.}(2020)\citenamefont {Foo}, \citenamefont {Onoe},\ and\ \citenamefont {Zych}}]{SuperTrajectories}%
  \BibitemOpen
  \bibfield  {author} {\bibinfo {author} {\bibfnamefont {J.}~\bibnamefont {Foo}}, \bibinfo {author} {\bibfnamefont {S.}~\bibnamefont {Onoe}},\ and\ \bibinfo {author} {\bibfnamefont {M.}~\bibnamefont {Zych}},\ }\bibfield  {title} {\bibinfo {title} {Unruh-{deWitt} detectors in quantum superpositions of trajectories},\ }\href {https://doi.org/10.1103/PhysRevD.102.085013} {\bibfield  {journal} {\bibinfo  {journal} {Phys. Rev. D}\ }\textbf {\bibinfo {volume} {102}},\ \bibinfo {pages} {085013} (\bibinfo {year} {2020})}\BibitemShut {NoStop}%
\bibitem [{\citenamefont {Barbado}\ \emph {et~al.}(2020)\citenamefont {Barbado}, \citenamefont {Castro-Ruiz}, \citenamefont {Apadula},\ and\ \citenamefont {Brukner}}]{UnruhSuperposition}%
  \BibitemOpen
  \bibfield  {author} {\bibinfo {author} {\bibfnamefont {L.~C.}\ \bibnamefont {Barbado}}, \bibinfo {author} {\bibfnamefont {E.}~\bibnamefont {Castro-Ruiz}}, \bibinfo {author} {\bibfnamefont {L.}~\bibnamefont {Apadula}},\ and\ \bibinfo {author} {\bibfnamefont {C.}~\bibnamefont {Brukner}},\ }\bibfield  {title} {\bibinfo {title} {Unruh effect for detectors in superposition of accelerations},\ }\href {https://doi.org/10.1103/PhysRevD.102.045002} {\bibfield  {journal} {\bibinfo  {journal} {Phys. Rev. D}\ }\textbf {\bibinfo {volume} {102}},\ \bibinfo {pages} {045002} (\bibinfo {year} {2020})}\BibitemShut {NoStop}%
\bibitem [{\citenamefont {Stritzelberger}\ and\ \citenamefont {Kempf}(2020)}]{NadineDelocalization}%
  \BibitemOpen
  \bibfield  {author} {\bibinfo {author} {\bibfnamefont {N.}~\bibnamefont {Stritzelberger}}\ and\ \bibinfo {author} {\bibfnamefont {A.}~\bibnamefont {Kempf}},\ }\bibfield  {title} {\bibinfo {title} {Coherent delocalization in the light-matter interaction},\ }\href {https://doi.org/10.1103/PhysRevD.101.036007} {\bibfield  {journal} {\bibinfo  {journal} {Phys. Rev. D}\ }\textbf {\bibinfo {volume} {101}},\ \bibinfo {pages} {036007} (\bibinfo {year} {2020})}\BibitemShut {NoStop}%
\bibitem [{\citenamefont {Giacomini}\ and\ \citenamefont {Kempf}(2022)}]{FlaminiaAchim}%
  \BibitemOpen
  \bibfield  {author} {\bibinfo {author} {\bibfnamefont {F.}~\bibnamefont {Giacomini}}\ and\ \bibinfo {author} {\bibfnamefont {A.}~\bibnamefont {Kempf}},\ }\bibfield  {title} {\bibinfo {title} {Second-quantized {Unruh-DeWitt} detectors and their quantum reference frame transformations},\ }\href {https://doi.org/10.1103/PhysRevD.105.125001} {\bibfield  {journal} {\bibinfo  {journal} {Phys. Rev. D}\ }\textbf {\bibinfo {volume} {105}},\ \bibinfo {pages} {125001} (\bibinfo {year} {2022})}\BibitemShut {NoStop}%
\bibitem [{\citenamefont {Gale}\ and\ \citenamefont {Zych}(2023)}]{EvanDelocalization}%
  \BibitemOpen
  \bibfield  {author} {\bibinfo {author} {\bibfnamefont {E.~P.~G.}\ \bibnamefont {Gale}}\ and\ \bibinfo {author} {\bibfnamefont {M.}~\bibnamefont {Zych}},\ }\bibfield  {title} {\bibinfo {title} {Relativistic {Unruh-DeWitt} detectors with quantized center of mass},\ }\href {https://doi.org/10.1103/PhysRevD.107.056023} {\bibfield  {journal} {\bibinfo  {journal} {Phys. Rev. D}\ }\textbf {\bibinfo {volume} {107}},\ \bibinfo {pages} {056023} (\bibinfo {year} {2023})}\BibitemShut {NoStop}%
\bibitem [{\citenamefont {Pozas-Kerstjens}\ and\ \citenamefont {Mart\'{i}n-Mart\'{i}nez}(2016)}]{Pozas2016}%
  \BibitemOpen
  \bibfield  {author} {\bibinfo {author} {\bibfnamefont {A.}~\bibnamefont {Pozas-Kerstjens}}\ and\ \bibinfo {author} {\bibfnamefont {E.}~\bibnamefont {Mart\'{i}n-Mart\'{i}nez}},\ }\bibfield  {title} {\bibinfo {title} {Entanglement harvesting from the electromagnetic vacuum with hydrogenlike atoms},\ }\href {https://doi.org/10.1103/PhysRevD.94.064074} {\bibfield  {journal} {\bibinfo  {journal} {Phys. Rev. D}\ }\textbf {\bibinfo {volume} {94}},\ \bibinfo {pages} {064074} (\bibinfo {year} {2016})}\BibitemShut {NoStop}%
\bibitem [{\citenamefont {Torres}\ \emph {et~al.}(2020)\citenamefont {Torres}, \citenamefont {Rick~Perche}, \citenamefont {Landulfo},\ and\ \citenamefont {Matsas}}]{neutrinos}%
  \BibitemOpen
  \bibfield  {author} {\bibinfo {author} {\bibfnamefont {B.~d. S.~L.}\ \bibnamefont {Torres}}, \bibinfo {author} {\bibfnamefont {T.}~\bibnamefont {Rick~Perche}}, \bibinfo {author} {\bibfnamefont {A.~G.~S.}\ \bibnamefont {Landulfo}},\ and\ \bibinfo {author} {\bibfnamefont {G.~E.~A.}\ \bibnamefont {Matsas}},\ }\bibfield  {title} {\bibinfo {title} {Neutrino flavor oscillations without flavor states},\ }\href {https://doi.org/10.1103/PhysRevD.102.093003} {\bibfield  {journal} {\bibinfo  {journal} {Phys. Rev. D}\ }\textbf {\bibinfo {volume} {102}},\ \bibinfo {pages} {093003} (\bibinfo {year} {2020})}\BibitemShut {NoStop}%
\bibitem [{\citenamefont {Perche}\ \emph {et~al.}(2022)\citenamefont {Perche}, \citenamefont {Ragula},\ and\ \citenamefont {Martín-Martínez}}]{boris}%
  \BibitemOpen
  \bibfield  {author} {\bibinfo {author} {\bibfnamefont {T.~R.}\ \bibnamefont {Perche}}, \bibinfo {author} {\bibfnamefont {B.}~\bibnamefont {Ragula}},\ and\ \bibinfo {author} {\bibfnamefont {E.}~\bibnamefont {Martín-Martínez}},\ }\href@noop {} {\bibinfo {title} {Harvesting entanglement from the gravitational vacuum}} (\bibinfo {year} {2022}),\ \Eprint {https://arxiv.org/abs/2210.14921} {arXiv:2210.14921 [quant-ph]} \BibitemShut {NoStop}%
\bibitem [{\citenamefont {Poisson}(2004)}]{poisson}%
  \BibitemOpen
  \bibfield  {author} {\bibinfo {author} {\bibfnamefont {E.}~\bibnamefont {Poisson}},\ }\bibfield  {title} {\bibinfo {title} {The motion of point particles in curved spacetime},\ }\bibfield  {journal} {\bibinfo  {journal} {Living Rev. Relativ.}\ }\textbf {\bibinfo {volume} {7}},\ \href {https://doi.org/10.12942/lrr-2004-6} {10.12942/lrr-2004-6} (\bibinfo {year} {2004})\BibitemShut {NoStop}%
\bibitem [{\citenamefont {Unruh}\ and\ \citenamefont {Wald}(1984)}]{Unruh-Wald}%
  \BibitemOpen
  \bibfield  {author} {\bibinfo {author} {\bibfnamefont {W.~G.}\ \bibnamefont {Unruh}}\ and\ \bibinfo {author} {\bibfnamefont {R.~M.}\ \bibnamefont {Wald}},\ }\bibfield  {title} {\bibinfo {title} {What happens when an accelerating observer detects a {R}indler particle},\ }\href {https://doi.org/10.1103/PhysRevD.29.1047} {\bibfield  {journal} {\bibinfo  {journal} {Phys. Rev. D}\ }\textbf {\bibinfo {volume} {29}},\ \bibinfo {pages} {1047} (\bibinfo {year} {1984})}\BibitemShut {NoStop}%
\bibitem [{\citenamefont {Louko}\ and\ \citenamefont {Satz}(2006)}]{JormaRigid}%
  \BibitemOpen
  \bibfield  {author} {\bibinfo {author} {\bibfnamefont {J.}~\bibnamefont {Louko}}\ and\ \bibinfo {author} {\bibfnamefont {A.}~\bibnamefont {Satz}},\ }\bibfield  {title} {\bibinfo {title} {How often does the {U}nruh-{DeWitt} detector click? regularization by a spatial profile},\ }\href {https://doi.org/10.1088/0264-9381/23/22/015} {\bibfield  {journal} {\bibinfo  {journal} {Class. Quantum Gravity}\ }\textbf {\bibinfo {volume} {23}},\ \bibinfo {pages} {6321} (\bibinfo {year} {2006})}\BibitemShut {NoStop}%
\bibitem [{\citenamefont {Louko}\ and\ \citenamefont {Satz}(2008)}]{LoukoCurvedSpacetimes}%
  \BibitemOpen
  \bibfield  {author} {\bibinfo {author} {\bibfnamefont {J.}~\bibnamefont {Louko}}\ and\ \bibinfo {author} {\bibfnamefont {A.}~\bibnamefont {Satz}},\ }\bibfield  {title} {\bibinfo {title} {Transition rate of the {U}nruh{\textendash}{DeWitt} detector in curved spacetime},\ }\href {https://doi.org/10.1088/0264-9381/25/5/055012} {\bibfield  {journal} {\bibinfo  {journal} {Class. Quantum Gravity}\ }\textbf {\bibinfo {volume} {25}},\ \bibinfo {pages} {055012} (\bibinfo {year} {2008})}\BibitemShut {NoStop}%
\bibitem [{\citenamefont {Crispino}\ \emph {et~al.}(2008)\citenamefont {Crispino}, \citenamefont {Higuchi},\ and\ \citenamefont {Matsas}}]{matsasUnruh}%
  \BibitemOpen
  \bibfield  {author} {\bibinfo {author} {\bibfnamefont {L.~C.~B.}\ \bibnamefont {Crispino}}, \bibinfo {author} {\bibfnamefont {A.}~\bibnamefont {Higuchi}},\ and\ \bibinfo {author} {\bibfnamefont {G.~E.~A.}\ \bibnamefont {Matsas}},\ }\bibfield  {title} {\bibinfo {title} {The {U}nruh effect and its applications},\ }\href {https://doi.org/10.1103/RevModPhys.80.787} {\bibfield  {journal} {\bibinfo  {journal} {Rev. Mod. Phys.}\ }\textbf {\bibinfo {volume} {80}},\ \bibinfo {pages} {787} (\bibinfo {year} {2008})}\BibitemShut {NoStop}%
\bibitem [{\citenamefont {Mart\'{\i}n-Mart\'{\i}nez}\ \emph {et~al.}(2013)\citenamefont {Mart\'{\i}n-Mart\'{\i}nez}, \citenamefont {Montero},\ and\ \citenamefont {del Rey}}]{eduardoOld}%
  \BibitemOpen
  \bibfield  {author} {\bibinfo {author} {\bibfnamefont {E.}~\bibnamefont {Mart\'{\i}n-Mart\'{\i}nez}}, \bibinfo {author} {\bibfnamefont {M.}~\bibnamefont {Montero}},\ and\ \bibinfo {author} {\bibfnamefont {M.}~\bibnamefont {del Rey}},\ }\bibfield  {title} {\bibinfo {title} {Wavepacket detection with the {Unruh-DeWitt} model},\ }\href {https://doi.org/10.1103/PhysRevD.87.064038} {\bibfield  {journal} {\bibinfo  {journal} {Phys. Rev. D}\ }\textbf {\bibinfo {volume} {87}},\ \bibinfo {pages} {064038} (\bibinfo {year} {2013})}\BibitemShut {NoStop}%
\bibitem [{\citenamefont {Simidzija}\ \emph {et~al.}(2020)\citenamefont {Simidzija}, \citenamefont {Ahmadzadegan}, \citenamefont {Kempf},\ and\ \citenamefont {Mart\'{i}n-Mart\'{i}nez}}]{Simidzija_2020}%
  \BibitemOpen
  \bibfield  {author} {\bibinfo {author} {\bibfnamefont {P.}~\bibnamefont {Simidzija}}, \bibinfo {author} {\bibfnamefont {A.}~\bibnamefont {Ahmadzadegan}}, \bibinfo {author} {\bibfnamefont {A.}~\bibnamefont {Kempf}},\ and\ \bibinfo {author} {\bibfnamefont {E.}~\bibnamefont {Mart\'{i}n-Mart\'{i}nez}},\ }\bibfield  {title} {\bibinfo {title} {Transmission of quantum information through quantum fields},\ }\href {https://doi.org/10.1103/PhysRevD.101.036014} {\bibfield  {journal} {\bibinfo  {journal} {Phys. Rev. D}\ }\textbf {\bibinfo {volume} {101}},\ \bibinfo {pages} {036014} (\bibinfo {year} {2020})}\BibitemShut {NoStop}%
\bibitem [{\citenamefont {Tjoa}\ and\ \citenamefont {Mann}(2020)}]{Tjoa2020}%
  \BibitemOpen
  \bibfield  {author} {\bibinfo {author} {\bibfnamefont {E.}~\bibnamefont {Tjoa}}\ and\ \bibinfo {author} {\bibfnamefont {R.~B.}\ \bibnamefont {Mann}},\ }\bibfield  {title} {\bibinfo {title} {Harvesting correlations in {S}chwarzschild and collapsing shell spacetimes},\ }\href {https://doi.org/10.1007/JHEP08(2020)155} {\bibfield  {journal} {\bibinfo  {journal} {J. High Energy Phys.}\ }\textbf {\bibinfo {volume} {2020}}\bibinfo  {number} { (8)},\ \bibinfo {pages} {155}}\BibitemShut {NoStop}%
\bibitem [{\citenamefont {Cozzella}\ \emph {et~al.}(2020)\citenamefont {Cozzella}, \citenamefont {Fulling}, \citenamefont {Landulfo},\ and\ \citenamefont {Matsas}}]{CozzellaUDWLimit}%
  \BibitemOpen
\bibfield  {number} {  }\bibfield  {author} {\bibinfo {author} {\bibfnamefont {G.}~\bibnamefont {Cozzella}}, \bibinfo {author} {\bibfnamefont {S.~A.}\ \bibnamefont {Fulling}}, \bibinfo {author} {\bibfnamefont {A.~G.~S.}\ \bibnamefont {Landulfo}},\ and\ \bibinfo {author} {\bibfnamefont {G.~E.~A.}\ \bibnamefont {Matsas}},\ }\bibfield  {title} {\bibinfo {title} {Uniformly accelerated classical sources as limits of {Unruh-DeWitt} detectors},\ }\href {https://doi.org/10.1103/PhysRevD.102.105016} {\bibfield  {journal} {\bibinfo  {journal} {Phys. Rev. D}\ }\textbf {\bibinfo {volume} {102}},\ \bibinfo {pages} {105016} (\bibinfo {year} {2020})}\BibitemShut {NoStop}%
\bibitem [{\citenamefont {Burbano}\ \emph {et~al.}(2021)\citenamefont {Burbano}, \citenamefont {Perche},\ and\ \citenamefont {de~S.~L.~Torres}}]{Ivan2021}%
  \BibitemOpen
  \bibfield  {author} {\bibinfo {author} {\bibfnamefont {I.~M.}\ \bibnamefont {Burbano}}, \bibinfo {author} {\bibfnamefont {T.~R.}\ \bibnamefont {Perche}},\ and\ \bibinfo {author} {\bibfnamefont {B.}~\bibnamefont {de~S.~L.~Torres}},\ }\bibfield  {title} {\bibinfo {title} {A path integral formulation for particle detectors: the {Unruh-DeWitt} model as a line defect},\ }\href {https://doi.org/10.1007/JHEP03(2021)076} {\bibfield  {journal} {\bibinfo  {journal} {J. High Energy Phys.}\ }\textbf {\bibinfo {volume} {2021}}\bibinfo  {number} { (3)},\ \bibinfo {pages} {76}}\BibitemShut {NoStop}%
\bibitem [{\citenamefont {Unruh}\ and\ \citenamefont {Zurek}(1989)}]{UnruhZurek}%
  \BibitemOpen
\bibfield  {number} {  }\bibfield  {author} {\bibinfo {author} {\bibfnamefont {W.~G.}\ \bibnamefont {Unruh}}\ and\ \bibinfo {author} {\bibfnamefont {W.~H.}\ \bibnamefont {Zurek}},\ }\bibfield  {title} {\bibinfo {title} {Reduction of a wave packet in quantum brownian motion},\ }\href {https://doi.org/10.1103/PhysRevD.40.1071} {\bibfield  {journal} {\bibinfo  {journal} {Phys. Rev. D}\ }\textbf {\bibinfo {volume} {40}},\ \bibinfo {pages} {1071} (\bibinfo {year} {1989})}\BibitemShut {NoStop}%
\bibitem [{\citenamefont {Hu}\ and\ \citenamefont {Matacz}(1994)}]{HuMatacs}%
  \BibitemOpen
  \bibfield  {author} {\bibinfo {author} {\bibfnamefont {B.~L.}\ \bibnamefont {Hu}}\ and\ \bibinfo {author} {\bibfnamefont {A.}~\bibnamefont {Matacz}},\ }\bibfield  {title} {\bibinfo {title} {Quantum brownian motion in a bath of parametric oscillators: A model for system-field interactions},\ }\href {https://doi.org/10.1103/PhysRevD.49.6612} {\bibfield  {journal} {\bibinfo  {journal} {Phys. Rev. D}\ }\textbf {\bibinfo {volume} {49}},\ \bibinfo {pages} {6612} (\bibinfo {year} {1994})}\BibitemShut {NoStop}%
\bibitem [{\citenamefont {Lin}\ and\ \citenamefont {Hu}(2007)}]{LinHu2007}%
  \BibitemOpen
  \bibfield  {author} {\bibinfo {author} {\bibfnamefont {S.-Y.}\ \bibnamefont {Lin}}\ and\ \bibinfo {author} {\bibfnamefont {B.~L.}\ \bibnamefont {Hu}},\ }\bibfield  {title} {\bibinfo {title} {Backreaction and the {U}nruh effect: New insights from exact solutions of uniformly accelerated detectors},\ }\href {https://doi.org/10.1103/PhysRevD.76.064008} {\bibfield  {journal} {\bibinfo  {journal} {Phys. Rev. D}\ }\textbf {\bibinfo {volume} {76}},\ \bibinfo {pages} {064008} (\bibinfo {year} {2007})}\BibitemShut {NoStop}%
\bibitem [{\citenamefont {Bruschi}\ \emph {et~al.}(2013)\citenamefont {Bruschi}, \citenamefont {Lee},\ and\ \citenamefont {Fuentes}}]{Bruschi_2013}%
  \BibitemOpen
  \bibfield  {author} {\bibinfo {author} {\bibfnamefont {D.~E.}\ \bibnamefont {Bruschi}}, \bibinfo {author} {\bibfnamefont {A.~R.}\ \bibnamefont {Lee}},\ and\ \bibinfo {author} {\bibfnamefont {I.}~\bibnamefont {Fuentes}},\ }\bibfield  {title} {\bibinfo {title} {Time evolution techniques for detectors in relativistic quantum information},\ }\href {https://doi.org/10.1088/1751-8113/46/16/165303} {\bibfield  {journal} {\bibinfo  {journal} {J. Phys. A: Math. Theor.}\ }\textbf {\bibinfo {volume} {46}},\ \bibinfo {pages} {165303} (\bibinfo {year} {2013})}\BibitemShut {NoStop}%
\bibitem [{\citenamefont {Brown}\ \emph {et~al.}(2013)\citenamefont {Brown}, \citenamefont {Mart\'{\i}n-Mart\'{\i}nez}, \citenamefont {Menicucci},\ and\ \citenamefont {Mann}}]{BrownHarmonic}%
  \BibitemOpen
  \bibfield  {author} {\bibinfo {author} {\bibfnamefont {E.~G.}\ \bibnamefont {Brown}}, \bibinfo {author} {\bibfnamefont {E.}~\bibnamefont {Mart\'{\i}n-Mart\'{\i}nez}}, \bibinfo {author} {\bibfnamefont {N.~C.}\ \bibnamefont {Menicucci}},\ and\ \bibinfo {author} {\bibfnamefont {R.~B.}\ \bibnamefont {Mann}},\ }\bibfield  {title} {\bibinfo {title} {Detectors for probing relativistic quantum physics beyond perturbation theory},\ }\href {https://doi.org/10.1103/PhysRevD.87.084062} {\bibfield  {journal} {\bibinfo  {journal} {Phys. Rev. D}\ }\textbf {\bibinfo {volume} {87}},\ \bibinfo {pages} {084062} (\bibinfo {year} {2013})}\BibitemShut {NoStop}%
\bibitem [{\citenamefont {Braunstein}\ and\ \citenamefont {van Loock}(2005)}]{ContinuousVariablesQI}%
  \BibitemOpen
  \bibfield  {author} {\bibinfo {author} {\bibfnamefont {S.~L.}\ \bibnamefont {Braunstein}}\ and\ \bibinfo {author} {\bibfnamefont {P.}~\bibnamefont {van Loock}},\ }\bibfield  {title} {\bibinfo {title} {Quantum information with continuous variables},\ }\href {https://doi.org/10.1103/RevModPhys.77.513} {\bibfield  {journal} {\bibinfo  {journal} {Rev. Mod. Phys.}\ }\textbf {\bibinfo {volume} {77}},\ \bibinfo {pages} {513} (\bibinfo {year} {2005})}\BibitemShut {NoStop}%
\bibitem [{\citenamefont {Weedbrook}\ \emph {et~al.}(2012)\citenamefont {Weedbrook}, \citenamefont {Pirandola}, \citenamefont {Garc\'{\i}a-Patr\'on}, \citenamefont {Cerf}, \citenamefont {Ralph}, \citenamefont {Shapiro},\ and\ \citenamefont {Lloyd}}]{gaussianquantuminfo}%
  \BibitemOpen
  \bibfield  {author} {\bibinfo {author} {\bibfnamefont {C.}~\bibnamefont {Weedbrook}}, \bibinfo {author} {\bibfnamefont {S.}~\bibnamefont {Pirandola}}, \bibinfo {author} {\bibfnamefont {R.}~\bibnamefont {Garc\'{\i}a-Patr\'on}}, \bibinfo {author} {\bibfnamefont {N.~J.}\ \bibnamefont {Cerf}}, \bibinfo {author} {\bibfnamefont {T.~C.}\ \bibnamefont {Ralph}}, \bibinfo {author} {\bibfnamefont {J.~H.}\ \bibnamefont {Shapiro}},\ and\ \bibinfo {author} {\bibfnamefont {S.}~\bibnamefont {Lloyd}},\ }\bibfield  {title} {\bibinfo {title} {Gaussian quantum information},\ }\href {https://doi.org/10.1103/RevModPhys.84.621} {\bibfield  {journal} {\bibinfo  {journal} {Rev. Mod. Phys.}\ }\textbf {\bibinfo {volume} {84}},\ \bibinfo {pages} {621} (\bibinfo {year} {2012})}\BibitemShut {NoStop}%
\bibitem [{\citenamefont {Adesso}\ \emph {et~al.}(2014)\citenamefont {Adesso}, \citenamefont {Ragy},\ and\ \citenamefont {Lee}}]{Adesso2014}%
  \BibitemOpen
  \bibfield  {author} {\bibinfo {author} {\bibfnamefont {G.}~\bibnamefont {Adesso}}, \bibinfo {author} {\bibfnamefont {S.}~\bibnamefont {Ragy}},\ and\ \bibinfo {author} {\bibfnamefont {A.~R.}\ \bibnamefont {Lee}},\ }\bibfield  {title} {\bibinfo {title} {Continuous variable quantum information: Gaussian states and beyond},\ }\href {https://doi.org/10.1142/S1230161214400010} {\bibfield  {journal} {\bibinfo  {journal} {Open Syst. Inf. Dyn.}\ }\textbf {\bibinfo {volume} {21}},\ \bibinfo {pages} {1440001} (\bibinfo {year} {2014})}\BibitemShut {NoStop}%
\bibitem [{\citenamefont {Hackl}\ and\ \citenamefont {Bianchi}(2021)}]{HacklKahler2021}%
  \BibitemOpen
  \bibfield  {author} {\bibinfo {author} {\bibfnamefont {L.}~\bibnamefont {Hackl}}\ and\ \bibinfo {author} {\bibfnamefont {E.}~\bibnamefont {Bianchi}},\ }\bibfield  {title} {\bibinfo {title} {{Bosonic and fermionic Gaussian states from Kähler structures}},\ }\href {https://doi.org/10.21468/SciPostPhysCore.4.3.025} {\bibfield  {journal} {\bibinfo  {journal} {SciPost Phys. Core}\ }\textbf {\bibinfo {volume} {4}},\ \bibinfo {pages} {025} (\bibinfo {year} {2021})}\BibitemShut {NoStop}%
\bibitem [{\citenamefont {Vriend}\ \emph {et~al.}(2021)\citenamefont {Vriend}, \citenamefont {Grimmer},\ and\ \citenamefont {Martín-Martínez}}]{SlowUnruh}%
  \BibitemOpen
  \bibfield  {author} {\bibinfo {author} {\bibfnamefont {S.}~\bibnamefont {Vriend}}, \bibinfo {author} {\bibfnamefont {D.}~\bibnamefont {Grimmer}},\ and\ \bibinfo {author} {\bibfnamefont {E.}~\bibnamefont {Martín-Martínez}},\ }\bibfield  {title} {\bibinfo {title} {The {U}nruh effect in slow motion},\ }\bibfield  {journal} {\bibinfo  {journal} {Symmetry}\ }\textbf {\bibinfo {volume} {13}},\ \href {https://doi.org/10.3390/sym13111977} {10.3390/sym13111977} (\bibinfo {year} {2021})\BibitemShut {NoStop}%
\bibitem [{\citenamefont {Mart\'{i}n-Mart\'{i}nez}\ and\ \citenamefont {Rodriguez-Lopez}(2018)}]{eduardo}%
  \BibitemOpen
  \bibfield  {author} {\bibinfo {author} {\bibfnamefont {E.}~\bibnamefont {Mart\'{i}n-Mart\'{i}nez}}\ and\ \bibinfo {author} {\bibfnamefont {P.}~\bibnamefont {Rodriguez-Lopez}},\ }\bibfield  {title} {\bibinfo {title} {Relativistic quantum optics: The relativistic invariance of the light-matter interaction models},\ }\href {https://doi.org/10.1103/PhysRevD.97.105026} {\bibfield  {journal} {\bibinfo  {journal} {Phys. Rev. D}\ }\textbf {\bibinfo {volume} {97}},\ \bibinfo {pages} {105026} (\bibinfo {year} {2018})}\BibitemShut {NoStop}%
\bibitem [{\citenamefont {Mart\'{\i}n-Mart\'{\i}nez}\ \emph {et~al.}(2020)\citenamefont {Mart\'{\i}n-Mart\'{\i}nez}, \citenamefont {Perche},\ and\ \citenamefont {de~S.~L.~Torres}}]{us}%
  \BibitemOpen
  \bibfield  {author} {\bibinfo {author} {\bibfnamefont {E.}~\bibnamefont {Mart\'{\i}n-Mart\'{\i}nez}}, \bibinfo {author} {\bibfnamefont {T.~R.}\ \bibnamefont {Perche}},\ and\ \bibinfo {author} {\bibfnamefont {B.}~\bibnamefont {de~S.~L.~Torres}},\ }\bibfield  {title} {\bibinfo {title} {General relativistic quantum optics: Finite-size particle detector models in curved spacetimes},\ }\href {https://doi.org/10.1103/PhysRevD.101.045017} {\bibfield  {journal} {\bibinfo  {journal} {Phys. Rev. D}\ }\textbf {\bibinfo {volume} {101}},\ \bibinfo {pages} {045017} (\bibinfo {year} {2020})}\BibitemShut {NoStop}%
\bibitem [{\citenamefont {Mart\'{\i}n-Mart\'{\i}nez}\ \emph {et~al.}(2021)\citenamefont {Mart\'{\i}n-Mart\'{\i}nez}, \citenamefont {Perche},\ and\ \citenamefont {Torres}}]{us2}%
  \BibitemOpen
  \bibfield  {author} {\bibinfo {author} {\bibfnamefont {E.}~\bibnamefont {Mart\'{\i}n-Mart\'{\i}nez}}, \bibinfo {author} {\bibfnamefont {T.~R.}\ \bibnamefont {Perche}},\ and\ \bibinfo {author} {\bibfnamefont {B.~d. S.~L.}\ \bibnamefont {Torres}},\ }\bibfield  {title} {\bibinfo {title} {Broken covariance of particle detector models in relativistic quantum information},\ }\href {https://doi.org/10.1103/PhysRevD.103.025007} {\bibfield  {journal} {\bibinfo  {journal} {Phys. Rev. D}\ }\textbf {\bibinfo {volume} {103}},\ \bibinfo {pages} {025007} (\bibinfo {year} {2021})}\BibitemShut {NoStop}%
\bibitem [{\citenamefont {Summers}\ and\ \citenamefont {Werner}(1985)}]{vacuumEntanglement}%
  \BibitemOpen
  \bibfield  {author} {\bibinfo {author} {\bibfnamefont {S.~J.}\ \bibnamefont {Summers}}\ and\ \bibinfo {author} {\bibfnamefont {R.}~\bibnamefont {Werner}},\ }\bibfield  {title} {\bibinfo {title} {The vacuum violates {B}ell's inequalities},\ }\href {https://doi.org/https://doi.org/10.1016/0375-9601(85)90093-3} {\bibfield  {journal} {\bibinfo  {journal} {Phys. lett., A}\ }\textbf {\bibinfo {volume} {110}},\ \bibinfo {pages} {257} (\bibinfo {year} {1985})}\BibitemShut {NoStop}%
\bibitem [{\citenamefont {Witten}(2018)}]{witten}%
  \BibitemOpen
  \bibfield  {author} {\bibinfo {author} {\bibfnamefont {E.}~\bibnamefont {Witten}},\ }\bibfield  {title} {\bibinfo {title} {{APS Medal for Exceptional Achievement in Research: Invited article on entanglement properties of quantum field theory}},\ }\href {https://doi.org/10.1103/RevModPhys.90.045003} {\bibfield  {journal} {\bibinfo  {journal} {Rev. Mod. Phys.}\ }\textbf {\bibinfo {volume} {90}},\ \bibinfo {pages} {045003} (\bibinfo {year} {2018})}\BibitemShut {NoStop}%
\bibitem [{\citenamefont {Reed}\ and\ \citenamefont {Simon}(1978)}]{ReedSimon4}%
  \BibitemOpen
  \bibfield  {author} {\bibinfo {author} {\bibfnamefont {M.}~\bibnamefont {Reed}}\ and\ \bibinfo {author} {\bibfnamefont {B.}~\bibnamefont {Simon}},\ }\href@noop {} {\emph {\bibinfo {title} {Methods of Modern Mathematical Physics, IV. Analysis of Operators}}}\ (\bibinfo  {publisher} {Academic Press},\ \bibinfo {year} {1978})\BibitemShut {NoStop}%
\bibitem [{\citenamefont {Simon}(2008)}]{Simon2008}%
  \BibitemOpen
  \bibfield  {author} {\bibinfo {author} {\bibfnamefont {B.}~\bibnamefont {Simon}},\ }\href@noop {} {\bibinfo {title} {Schr\"{o}dinger operators with purely discrete spectrum}} (\bibinfo {year} {2008}),\ \Eprint {https://arxiv.org/abs/0810.3275} {arXiv:0810.3275 [math.SP]} \BibitemShut {NoStop}%
\bibitem [{\citenamefont {Burbano}\ and\ \citenamefont {Calderón}(2021)}]{IvanPathIntegrals}%
  \BibitemOpen
  \bibfield  {author} {\bibinfo {author} {\bibfnamefont {I.~M.}\ \bibnamefont {Burbano}}\ and\ \bibinfo {author} {\bibfnamefont {F.}~\bibnamefont {Calderón}},\ }\href@noop {} {\bibinfo {title} {Self-normalizing path integrals}} (\bibinfo {year} {2021}),\ \Eprint {https://arxiv.org/abs/2109.00517} {arXiv:2109.00517 [hep-th]} \BibitemShut {NoStop}%
\bibitem [{\citenamefont {BenTov}(2021)}]{BenTov}%
  \BibitemOpen
  \bibfield  {author} {\bibinfo {author} {\bibfnamefont {Y.}~\bibnamefont {BenTov}},\ }\href@noop {} {\bibinfo {title} {Schwinger-{K}eldysh path integral for the quantum harmonic oscillator}} (\bibinfo {year} {2021}),\ \Eprint {https://arxiv.org/abs/2102.05029} {arXiv:2102.05029 [hep-th]} \BibitemShut {NoStop}%
\bibitem [{\citenamefont {Rammer}(2007)}]{rammer_2007}%
  \BibitemOpen
  \bibfield  {author} {\bibinfo {author} {\bibfnamefont {J.}~\bibnamefont {Rammer}},\ }\href {https://doi.org/10.1017/CBO9780511618956} {\emph {\bibinfo {title} {Quantum Field Theory of Non-equilibrium States}}}\ (\bibinfo  {publisher} {Cambridge University Press},\ \bibinfo {year} {2007})\BibitemShut {NoStop}%
\bibitem [{\citenamefont {Calzetta}\ and\ \citenamefont {Hu}(2008)}]{calzetta_hu_2008}%
  \BibitemOpen
  \bibfield  {author} {\bibinfo {author} {\bibfnamefont {E.~A.}\ \bibnamefont {Calzetta}}\ and\ \bibinfo {author} {\bibfnamefont {B.-L.}\ \bibnamefont {Hu}},\ }\href {https://doi.org/10.1017/CBO9780511535123} {\emph {\bibinfo {title} {Nonequilibrium Quantum Field Theory}}},\ Cambridge Monographs on Mathematical Physics\ (\bibinfo  {publisher} {Cambridge University Press},\ \bibinfo {year} {2008})\BibitemShut {NoStop}%
\bibitem [{\citenamefont {Dowker}\ and\ \citenamefont {Halliwell}(1992)}]{DecoherenceFunctional}%
  \BibitemOpen
  \bibfield  {author} {\bibinfo {author} {\bibfnamefont {H.~F.}\ \bibnamefont {Dowker}}\ and\ \bibinfo {author} {\bibfnamefont {J.~J.}\ \bibnamefont {Halliwell}},\ }\bibfield  {title} {\bibinfo {title} {Quantum mechanics of history: The decoherence functional in quantum mechanics},\ }\href {https://doi.org/10.1103/PhysRevD.46.1580} {\bibfield  {journal} {\bibinfo  {journal} {Phys. Rev. D}\ }\textbf {\bibinfo {volume} {46}},\ \bibinfo {pages} {1580} (\bibinfo {year} {1992})}\BibitemShut {NoStop}%
\bibitem [{\citenamefont {Calzetta}\ and\ \citenamefont {Hu}(1987)}]{CalzettaClosed}%
  \BibitemOpen
  \bibfield  {author} {\bibinfo {author} {\bibfnamefont {E.}~\bibnamefont {Calzetta}}\ and\ \bibinfo {author} {\bibfnamefont {B.~L.}\ \bibnamefont {Hu}},\ }\bibfield  {title} {\bibinfo {title} {Closed-time-path functional formalism in curved spacetime: Application to cosmological back-reaction problems},\ }\href {https://doi.org/10.1103/PhysRevD.35.495} {\bibfield  {journal} {\bibinfo  {journal} {Phys. Rev. D}\ }\textbf {\bibinfo {volume} {35}},\ \bibinfo {pages} {495} (\bibinfo {year} {1987})}\BibitemShut {NoStop}%
\bibitem [{\citenamefont {Adshead}\ \emph {et~al.}(2009)\citenamefont {Adshead}, \citenamefont {Easther},\ and\ \citenamefont {Lim}}]{InInFormalism}%
  \BibitemOpen
  \bibfield  {author} {\bibinfo {author} {\bibfnamefont {P.}~\bibnamefont {Adshead}}, \bibinfo {author} {\bibfnamefont {R.}~\bibnamefont {Easther}},\ and\ \bibinfo {author} {\bibfnamefont {E.~A.}\ \bibnamefont {Lim}},\ }\bibfield  {title} {\bibinfo {title} {``{I}n-in'' formalism and cosmological perturbations},\ }\href {https://doi.org/10.1103/PhysRevD.80.083521} {\bibfield  {journal} {\bibinfo  {journal} {Phys. Rev. D}\ }\textbf {\bibinfo {volume} {80}},\ \bibinfo {pages} {083521} (\bibinfo {year} {2009})}\BibitemShut {NoStop}%
\bibitem [{\citenamefont {Ruep}(2021)}]{max}%
  \BibitemOpen
  \bibfield  {author} {\bibinfo {author} {\bibfnamefont {M.~H.}\ \bibnamefont {Ruep}},\ }\bibfield  {title} {\bibinfo {title} {Weakly coupled local particle detectors cannot harvest entanglement},\ }\href {https://doi.org/10.1088/1361-6382/ac1b08} {\bibfield  {journal} {\bibinfo  {journal} {Class. Quantum Gravity}\ }\textbf {\bibinfo {volume} {38}},\ \bibinfo {pages} {195029} (\bibinfo {year} {2021})}\BibitemShut {NoStop}%
\bibitem [{\citenamefont {Grimmer}\ \emph {et~al.}(2021)\citenamefont {Grimmer}, \citenamefont {Torres},\ and\ \citenamefont {Mart\'{\i}n-Mart\'{\i}nez}}]{RuepReply}%
  \BibitemOpen
  \bibfield  {author} {\bibinfo {author} {\bibfnamefont {D.}~\bibnamefont {Grimmer}}, \bibinfo {author} {\bibfnamefont {B.~d. S.~L.}\ \bibnamefont {Torres}},\ and\ \bibinfo {author} {\bibfnamefont {E.}~\bibnamefont {Mart\'{\i}n-Mart\'{\i}nez}},\ }\bibfield  {title} {\bibinfo {title} {Measurements in {QFT}: Weakly coupled local particle detectors and entanglement harvesting},\ }\href {https://doi.org/10.1103/PhysRevD.104.085014} {\bibfield  {journal} {\bibinfo  {journal} {Phys. Rev. D}\ }\textbf {\bibinfo {volume} {104}},\ \bibinfo {pages} {085014} (\bibinfo {year} {2021})}\BibitemShut {NoStop}%
\bibitem [{\citenamefont {Perche}\ \emph {et~al.}(2024{\natexlab{b}})\citenamefont {Perche}, \citenamefont {Polo-G\'omez}, \citenamefont {Torres},\ and\ \citenamefont {Mart\'{\i}n-Mart\'{\i}nez}}]{QFTPDHarvesting}%
  \BibitemOpen
  \bibfield  {author} {\bibinfo {author} {\bibfnamefont {T.~R.}\ \bibnamefont {Perche}}, \bibinfo {author} {\bibfnamefont {J.}~\bibnamefont {Polo-G\'omez}}, \bibinfo {author} {\bibfnamefont {B.~d. S.~L.}\ \bibnamefont {Torres}},\ and\ \bibinfo {author} {\bibfnamefont {E.}~\bibnamefont {Mart\'{\i}n-Mart\'{\i}nez}},\ }\bibfield  {title} {\bibinfo {title} {Fully relativistic entanglement harvesting},\ }\href {https://doi.org/10.1103/PhysRevD.109.045018} {\bibfield  {journal} {\bibinfo  {journal} {Phys. Rev. D}\ }\textbf {\bibinfo {volume} {109}},\ \bibinfo {pages} {045018} (\bibinfo {year} {2024}{\natexlab{b}})}\BibitemShut {NoStop}%
\bibitem [{\citenamefont {de~Aguiar~Alves}(2023)}]{Alves2023}%
  \BibitemOpen
  \bibfield  {author} {\bibinfo {author} {\bibfnamefont {N.}~\bibnamefont {de~Aguiar~Alves}},\ }\href@noop {} {\bibinfo {title} {Nonperturbative aspects of quantum field theory in curved spacetime}} (\bibinfo {year} {2023}),\ \Eprint {https://arxiv.org/abs/2305.17453} {arXiv:2305.17453 [gr-qc]} \BibitemShut {NoStop}%
\end{thebibliography}%

\end{document}